\documentclass[english]{emulateapj}

\usepackage{textcomp}
\usepackage{amsmath,amssymb,amsfonts}
\usepackage{hyperref}
\usepackage{txfonts}

\usepackage{sidecap}
\usepackage{natbib}

\usepackage{booktabs} 
\usepackage[normalem]{ulem} 
\usepackage[table]{xcolor} 
\usepackage[FIGTOPCAP,Large]{subfigure}

\usepackage{aas_macros} 

\begin{document}

\title{\bf{Realistic outcomes of moon-moon collisions in Lunar formation theory}}

\author{Uri Malamud{$^1$}, Hagai B. Perets{$^{1,2}$}}

\affil{{$^1$}Department of Physics, Technion Israel Institute of Technology, Israel}
\affil{{$^2$}Department of Natural Sciences, The Open University of Israel, 1 University Road, PO Box 808, Raanana 4353701, Israel}

\email{urimala@physics.technion.ac.il}

\begin{abstract}
The multiple impact hypothesis proposes that the Moon formed through a series of smaller collisions, rather than a single giant impact. This study advances our understanding of this hypothesis, as well as moon collisions in other contexts, by exploring the implications of these smaller impacts, employing a novel methodological approach that combines self-consistent initial conditions, hybrid hydrodynamic/N-body simulations, and the incorporation of material strength.
Our findings challenge the conventional assumption of perfect mergers in previous models, revealing a spectrum of collision outcomes including partial accretion and mass loss. These outcomes are sensitive to collision parameters and Earth's tidal influence, underscoring the complex dynamics of lunar accretion. Importantly, we demonstrate that incorporating material strength is important for accurately simulating moonlet-sized impacts. This inclusion significantly affects fragmentation, tidal disruption, and the amount of material ejected or accreted onto Earth, ultimately impacting the Moon's growth trajectory. By accurately modeling diverse collision outcomes, our hybrid approach provides a powerful new framework for understanding the Moon's formation. We show that most collisions ($\approx90\%$) do not significantly erode the largest moonlet, supporting the feasibility of lunar growth through accretion. Moreover, we revise previous estimates of satellite disruption, suggesting a higher survival rate and further bolstering the multiple-impact scenario.  
\end{abstract}

\section{Introduction}\label{S:Intro}
Terrestrial planets are thought to form during the first 10-100 Myr of the evolution of a planetary system. During the last stages of terrestrial planet formation, planets grow mainly through many impacts with large planetary embryos (e.g. \cite{Agn+99}). In the most studied scenario, the giant impact scenario, the Moon is thought to have formed following the last (and single) such giant impact to occur with the proto-Earth. In this model the Moon typically coagulates from a debris disk formed following this impact; and is affected by the mutual Earth-Moon tides, leading to its outward migration, thus explaining its current location. However, the (single) giant-impact scenario, though extensively studied, still encounters major challenges in explaining the Earth-Moon system and its physical and compositional properties (see \cite{mas+15,Mas+17}, and recent reviews by \cite{Asp+14,Can+23} and references therein). More importantly, such a scenario is intrinsically incomplete and disconnected from the wider picture of terrestrial planet formation, in which the proto-Earth had experienced and grown through multiple giant impacts \citep{Agn+99}. Consideration of only the last giant impact in the prevailing paradigm disregards the critical evolution taking place prior to (and possibly following) this event. In particular, it ignores smaller-scale impacts which are deemed incapable of forming large moons in a single event. Furthermore, even in the context of the single giant-impact hypothesis, models that try to overcome the aforementioned difficulties such as high angular momentum impacts \citep{CukStewart-2012} or Synestia \cite{loc+18} require specific initial conditions for the Earth, which are nevertheless dictated by its previous impact history. Indeed, an initial fast spin of the Earth is required for high angular momentum impacts or the Synestia models, but it could be challenged by a history of multiple impacts because then a fast rotation of the Earth is difficult to achieve \citep{RufuEtAl-2017,Can+23}. Similarly, solutions requiring a pre-existing Earth magma ocean \citep{Hos+19} depend on previous impacts and their resulting heating of the Earth \citep[e.g.][]{Cit+22}.  In addition, the probability for a late high mass-ratio giant impact as required in the giant impact hypothesis, is by itself low ($~6.7\%$; \cite{Mas+17}), further making the scenario more fine-tuned. Finally, multiple moons might in principle form following the single impact scenario \citep{Can+99}, and therefore understanding the interactions and collisions of moons in the framework of multiple impacts, and even in the context of the single giant-impact scenario, is important.

Our working hypothesis, following the current understanding of terrestrial planet formation, is that such small impacts are the norm during planet formation. They are significantly more abundant could form smaller moons, and play a major role in the evolution of the proto-Earth. In previous studies, we have suggested that multiple moons have formed around the proto-Earth following multiple planetary-scale impacts in the past, a sequence that has thereby affected the evolution of both the Earth and the Moon \citep{Rin89,Cit+14,RufuEtAl-2017,CitronEtAl-2018,MalamudEtAl-2018}.

In the multiple impact framework, the moon formation processes described above occur repeatedly after each planetary-scale impact on the proto-Earth. A moonlet (or even several moonlets) formed in a previous impact on the proto-Earth, may already exist when another external proto-planet perturber interacts with the system. There are several potential modes of interaction, one of which is to form a new debris disk, and eventually another moonlet (as a result of coagulation from the disk). The new moonlet could then migrate outwards (due to tides) and eventually interact with a pre-existing (by then slower-to-migrate) moonlet. Possible outcomes of this interaction \citep{CitronEtAl-2018} are a collision between the two moonlets, or ejection/infall onto the proto-Earth (typically one of the moonlets, see \cite{MalamudEtAl-2018}). Cases in which both moonlets did not survive are referred to as disruptive. \cite{CitronEtAl-2018} tried to estimate the relative probabilities of each of these outcomes. They found them to depend on various parameters (especially the mass ratio between the outer and inner moon), however, if one were to generalize their results, the probability for a merger was found to be approximately similar to that of survival of only the outer moon (which was assumed to be more massive). The probability of completely disrupting the system was found to be somewhat smaller but still of the same order of magnitude. This meant that in the framework of multiple impacts, each impact had a reasonable chance to increase the moon's mass or at least keep the system from disrupting.

However, in the framework of the \cite{CitronEtAl-2018} study, as typically in other dynamical studies \citep{HaghighipourRaymond-2007,FischerCiesla-2014}, every merger was assumed to be a perfectly plastic merger. In reality, a physical collision between two moonlets does not necessarily imply a convergence of their masses. Hence, the aforementioned merger probability should be revised by considering a more realistic collision outcome, expecting it to branch into several different probabilities. These depend primarily on the collision velocity \citep{Asphaug-2010}. A perfect merger is more likely to occur at low velocities, while higher velocities may lead to hit-and-run outcomes or even a partially/fully disruptive collision in which a debris disk forms. The mass of the debris (e.g, relative to that of the largest fragment) increases with the collision velocity \citep{GendaEtAl-2012,LeinhardtStewart-2012,EmsenhuberEtAl-2024}. Our goal in this study is to track the realm of potential outcomes in moon-moon collisions by combining two different simulations for the short- and long-term collision evolution, thereby revisiting the actual sequence of events subsequent to a collision, and in turn, revising the merger probabilities that were found in previous studies.

The effects of the Earth during moon-moon collisions are not negligible since the planet's tidal forces can influence the distribution of material during a merger \citep{RufuAharonson-2019}. During more disruptive collisions, the escape velocity from the Earth (below about 10-15 R$_\oplus$) is considerably larger than the Lunar escape velocity, hence increasing the likelihood of trapping fragments in bound Earth orbit. Additionally, as major moonlets or other minor fragments in the debris field of a disruptive collision enter the Roche limit, they can be tidally disrupted. One previous study has looked into the collision dynamics of moonlets within the Earth's potential \citep{RufuAharonson-2019}. However, here we aim to introduce several methodological improvements: (a) We will perform hydrodynamic collision simulations considering material strength (instead of pure hydrodynamic simulations). As recently demonstrated by \cite{EmsenhuberEtAl-2024} via a large ensemble of giant impact collisions with material strength, the latter is important for collisions of moonlet-sized objects (even without a central planet as proposed here, and certainly when tidal effects by a planet are involved); (b) We will adopt self-consistent collision initial conditions directly from the N-body simulations of \cite{CitronEtAl-2018}, instead of performing a general parametric study where otherwise ad-hoc values are chosen for the velocity, impact angle and distance from the Earth; (c) We will alternate back and forth between short time scale hydrodynamic impact simulations and much longer time scale N-body simulations, tracking a complete and detailed chain of events leading to the collision outcome (instead of only simulating the initial collision and making various guesses regarding the subsequent evolution).

\section{Methodology}\label{S:Methods}
\subsection{Initial setup}\label{SS:InitialSetup}
In order to account for moonlet collisions in a self-consistent manner, we rely on direct results obtained by a previous study performed by \cite{CitronEtAl-2018}. These authors considered the gravitational and tidal interactions among past moonlets of the Earth. Their study, in turn, relied on the previous results of \cite{RufuEtAl-2017} which studied the impacts of the proto-Earth that led to the formation of debris disks, giving rise to moonlets whose mass could be estimated.

\cite{CitronEtAl-2018} modeled a two-moonlet system with one outer moonlet that has tidally evolved outward over millions of years, before a subsequent giant impact formed a new, inner moonlet, which was assumed to form slightly outside the fluid Roche limit by coagulating from a disk of debris. The dynamical evolution of such a system was examined for a variety of initial conditions to determine the probability of moonlet mergers or other dynamic outcomes. They used the \textit{Mercury-T} N-body code \citep{BolmontEtAl-2015}, which incorporates a prescription to compute tidal evolution. They considered a range of moonlet masses between 0.1 and 0.8 M$_{\rm L}$ (where M$_{\rm L}$ is the Lunar mass), outer moonlet semi-major axes between 10 and 30 R$_{\oplus}$ (where R$_{\oplus}$ is the Earth's radius) and otherwise random orbital elements besides the initial eccentricity which was assumed to be zero to account for eccentricity damping. In total, \cite{CitronEtAl-2018} performed more than 1000 simulations.

The \cite{CitronEtAl-2018} simulations resulted in a large set of moon-moon collisions, amounting to a few hundred cases overall. Since execution/analysis of our detailed follow-up simulations here, are both time-consuming and computationally expensive, we selected only 20 collisions from their entire data set as representative cases. In order to choose intelligently, we binned the original \cite{CitronEtAl-2018} data according to impact angle $\alpha$, and in the range between 0$^\circ$ (head-on) and 90$^\circ$ (fully grazing), we chose our 20 cases such that impact angles were separated by about 4-5$^\circ$. In addition, we also made sure to select a wide range of values for the impact velocities (1125-5215 m/s, or approximately 0.7-2.75 of the respective mutual escape velocities), moonlet masses (range between 0.11-0.78 M$_{\rm L}$; final total masses (range between 0.25-0.94 M$_{\rm L}$); ratios of moonlet masses (range between 1-5) and collision distances from the Earth (3.64-26.6 $R_\oplus$, or approximately 1.25-9 the fluid Roche radius $R_{\rm Roche\_f}$). Our explicit choice for these parameters is specified in Table \ref{tab:Cases}.

\begin{table*}
	\centering
	\begin{tabular}{|l|c|c|c|c|c|c|c|c|c|}
		\hline
		\textbf{Label}$^{\rm a}$ & $m_1$ ($M_{\rm L}$) & $a_1$ ($R_\oplus$) & $e_1$ & $m_2$ ($M_{\rm L}$) & $a_2$ ($R_\oplus$) & $e_2$ & $d$ ($R_\oplus$) & $v$ (m/s) & $\theta$ ($^\circ$) \\ \hline
		
		21b\_59 & 0.2754 & 23.5303 & 0.7138 & 0.3987 & 5.4280 &	0.2446 & 6.7574 & 5211 & 5.1741 \\
  
		21u\_23 & 0.1899 & 19.4398 & 0.3775 & 0.4712 & -44.9999$^{\rm b}$ &	1.5437 & 26.6062 & 1570 & 7.5219 \\
  
		20n\_53 & 0.3034 & 6.7005 & 0.1742 & 0.2622 & 21.8738 & 0.7112 & 7.8822 & 4223 & 12.3487 \\

        20f\_143 & 0.2282 & 3.5836 & 0.0168 & 0.1525 & 102.8175 & 0.9670 & 3.6471 & 5215 & 12.8393 \\
  
		20c\_35 & 0.1204 & 8.3352 & 0.3625 & 0.2039 & 6.1103 & 0.2877 & 5.3657 & 3481 & 18.7456 \\
  
		21f\_148 & 0.1797 & 17.3381 & 0.3790 & 0.3098 & 114.5991 & 0.8393 & 20.0726 & 1474 & 23.5615 \\
  
		20f\_118 & 0.2134 & 10.0396 & 0.6896 & 0.1454 & 7.7981 & 0.1787 & 8.8803 & 2769 & 27.5137 \\
  
		20\_2 & 0.4405 & 6.0024 & 0.3656 & 0.2747 & 44.9625 & 0.8281 & 7.7528 & 4572 & 28.2205 \\
  
		20f\_129 & 0.3249 & 8.9310 & 0.1419 & 0.2471 & 17.6731 & 0.7873 & 7.9178 & 2895 & 31.8609 \\
  
		21m\_53 & 0.1401 & 22.9960 & 0.1985 & 0.3546 & 19.7121 & 0.5444 & 19.8411 & 1344 & 35.9289 \\
  
		20f\_101 & 0.4342 & 6.6934 & 0.0968 & 0.1927 & 6.2265 & 0.5068 & 6.5583 & 2017 & 40.5662 \\
  
		20n\_91 & 0.3527 & 5.5171 & 0.0345 & 0.1467 & 9.8422 & 0.4245 & 5.7110 & 2920 & 48.1775 \\
  
		20p\_17 & 0.111 & 41.3865 & 0.5502 & 0.1169 & 14.3869 & 0.3688 & 19.0135 & 1125 & 55.1198 \\
  
		20\_166 & 0.2265 & 5.6437 & 0.2250 & 0.1988 & 14.8750 & 0.5939 & 6.4252 & 4361 & 57.2882 \\
  
		21\_215 & 0.2208 & 22.6988 & 0.6243 & 0.3280 & 7.5019 & 0.1638 & 8.6137 & 3828 & 65.4285 \\

        21c\_201 & 0.264 & 20.9548 & 0.4227 & 0.3155 & 9.415 & 0.6117 & 12.2036 & 1847 & 69.5652 \\
        
        22\_174 & 0.1523 & 30.4142 & 0.4320 & 0.7809 & 13.4800 & 0.5933 & 18.4094 & 3387 & 71.3929 \\

        21\_32 & 0.2662 & 11.8651 & 0.1969 & 0.3269 & 48.9187 & 0.7478 & 12.3241 & 3087 & 79.1707 \\

        20c\_46 & 0.3013 & 6.6675 & 0.2520 & 0.1763 & 16.9452 & 0.7411 & 7.6556 & 4077 & 84.1095 \\

        20e\_19 & 0.2288 & 13.5966 & 0.4688 & 0.3628 & 54.8004 & 0.6723 & 18.7428 & 1454 & 86.9678 \\
		\hline		
	\end{tabular}
    \caption{Collision initial conditions. Columns show (in order): label; mass ($m_1$) semi-major axis ($a_1$) and eccentricity ($e_1$) of moonlet 1; mass ($m_2$) semi-major axis ($a_2$) and eccentricity ($e_2$) of moonlet 2; distance from the Earth during impact ($d$); velocity of impact ($v$); angle of impact ($\theta$).
    \newline $^{\rm a}$ label names were adopted based on the classification of \cite{CitronEtAl-2018} for consistency.
    \newline $^{\rm b}$ Negative $a$ corresponds to $e>1$ and ensues from rare dynamic interactions which led to a hyperbolic orbit in one of the moonlets.}
	\label{tab:Cases}
\end{table*}

\subsection{Hydrodynamical simulations}\label{SS:Hydro}
We perform hydrodynamical collision simulations using the SPH impact code \textit{miluphcuda} \citep{SchaferEtAl-2020}. The code is implemented via CUDA, and runs on graphic processing units (GPU), with a substantial improvement in computation time compared to ordinary CPUs. The code implements a Barnes-Hut tree that allows for the treatment of self-gravity, as well as gas, fluid, elastic, and
plastic solid bodies, including a failure model for brittle materials and sub-resolution porosity treatment. Here we neglect porosity (justified by the size and thermal state of typical moonlets) and perform our simulations both with solid material strength (default model presented below), as well as in pure hydrodynamic mode, comparing these two cases. We use the M-ANEOS equation of state for dunite, assuming a single material for the moonlets, and neglect the relatively small Lunar iron fraction \citep{WeberEtAl-2011,BriaudEtAl-2023}. Our M-ANEOS dunite parameter input files are derived from \cite{Melosh-2007}. In order to treat numerical rotational instabilities, a tensorial correction scheme \citep{Speith-2006} is implemented.

For setting up each collision, we use the moonlet position and velocity coordinates immediately before the impact, as obtained from the \cite{CitronEtAl-2018} data. Given the moonlet masses, and that of the Earth, we can set up the collision using a pre-processing step. We assume that both impactor and target moonlets are non-rotating prior to impact, since although including rotation as a free parameter is technically possible, it would entail a very large increase in the number of simulations. Both impactor and target are generated as spheres, with relaxed internal structures, i.e. having hydrostatic density profiles and internal energy values from adiabatic compression, following the algorithm provided in appendix A of \cite{BurgerEtAl-2018}. This self-consistent semi-analytical calculation (i.e., using the same constituent physical relations as in the SPH model) replaces the otherwise necessary and far slower process of simulating the relaxation of each body in isolation. However, since here we are dealing with very small moons rather than planets, the density profile is almost uniform, and, more importantly, the internal heat ensuing from this calculation is small, resulting in cold temperatures up to a few hundred K. This cold start is not necessarily realistic since the cooling time between moonlet formation and subsequent interaction may be too small to reach a cool state, and therefore both the previous and the newer moonlets may be hotter in reality. Elevated temperatures reduce the effects of material strength, as elaborated on further below, and therefore the cold start setup considered here is really expected to maximize the importance of the material strength model, and significantly contrast with the purely hydro simulations. 

The fiducial resolution for our SPH simulations (either with material strength or pure hydro) is 5 $\times$ 10$^5$ particles, comprising the two moonlets. In a few of the most disruptive impacts, we have used a lower resolution of 10$^5$ particles, because disruptive simulations generate a lot of debris and are much more computationally expensive. In addition, the large number of debris fragments in highly disruptive collisions makes the subsequent N-body simulations much more computationally expensive. Taken together, lower-resolution disruptive SPH simulations are more manageable. We have conducted convergence tests, showing that relatively lower-resolution SPH outcomes are sufficiently similar to those of the highest resolution (Appendix \ref{Appendix:A}). 

Unlike the moonlets, the Earth is represented in the hydrodynamic simulations by a single gravitating SPH particle. This particle interacts with other SPH particles only through gravity. When other particles enter its physical cross-section (the Earth's radius), they are removed from the simulation, contributing their masses. This prevents the need for the alternative solution of using different particle resolutions when setting up the moonlets and the Earth, which can cause known SPH discontinuity issues.

For the plasticity model, we utilize a pressure-dependent yield strength following \cite{CollinsEtAl-2004} and using a similar implementation to \cite{Jutzi-2015}. Here the yield stress $Y_i$ is given by

\begin{equation}
Y_i=Y_0+\frac{\mu P}{1+\mu P / (Y_M-Y_0)},
\label{eq:YieldStress}
\end{equation}

where $Y_0$ is the cohesion (we use 10 MPa from \cite{OgawaEtAl-2021}), $\mu$ is the coefficient of friction (we use 0.7 from \cite{JutziAsphaug-2011}) and $Y_M$ is the shear strength at $P=\infty$ (we use 3.5 GPa from \cite{OgawaEtAl-2021,EmsenhuberEtAl-2024}). Additionally, the yield stress depends on temperature and drops to zero at the melting point. This behavior is typically approximated in various studies such as \cite{BenzAsphaug-1999,CollinsEtAl-2004,Jutzi-2015} and others, by a prescription that is considered viable for decreasing the elastic limit for increasing temperatures

\begin{equation}
Y_i \leftarrow Y_i \left( 1 - \frac{e}{e_{\rm melt}} \right),
\label{eq:YieldStressTemperatureDependence}
\end{equation}

where $e$ is the specific internal energy and $e_{\rm melt}$ is the specific melting energy (we use 3.4$\times$10$^6$ J/Kg from \cite{BenzAsphaug-1999}). Due to the complexity of using multiple parameters for distinct model realizations, we have decided not to vary any of the aforementioned plasticity model parameters, which should be kept in mind as a caveat of the model. We run our fiducial simulations for 28 h, which we find to be ample time in most cases since we usually find the most significant interactions to occur within the first several hours. After we conclude the hydrodynamic simulations, we analyze the outcomes, and unless there is a perfect merger, we hand over the data to a subsequent N-body simulation (as described next in Section \ref{SS:NBody}).

We note that the long-term gravitational dynamics are more accurate in our N-body code, since in the \textit{miluphcuda} SPH code we utilize the Barnes-Hut tree implementation for self-gravity. This considerably improves computational performance, compared to a direct calculation of the gravitational forces among numerous SPH particles. A direct computation is possible, but very slow given our resolution, which is why it is typically avoided in our as well as in other SPH studies. In order to verify that the N-body code is more accurate, we run a unique comparison case, 20n\_91, where the N-body simulation indicated that a second re-collision occurred quickly, after merely 5.5 days (compared to more typical times of weeks to years). We run the corresponding hydrodynamic simulation for the same duration of 5.5 days, which is computationally taxing but still possible. We found that in the SPH simulation the second collision did not occur, given the reduced accuracy.

\subsection{N-body simulations}\label{SS:NBody}
Although low-velocity impacts typically result in a merger into a single moonlet, high-velocity and/or grazing angles can lead to other outcomes. In order to study the longer-term dynamics of either two surviving moonlets with/out debris, or the full coagulation of debris fragments (for disruptive collisions), we introduce an N-body follow-up calculation in these cases.

The outcomes ensuing from SPH simulations are handed over to the open-source N-body code \textit{REBOUND}, via a special tool which we have developed previously in \cite{MalamudEtAl-2018}, and updated here to incorporate the latest release of \textit{REBOUND} (version 4). The hand-off tool initially reads our SPH output files which are in turn synthesized based on an analysis that finds physical fragments (clumps) of spatially connected SPH particles using a friends-of-friends algorithm. Fragment data is then passed on as recognizable input particles, to a modified \textit{REBOUND} code. The N-body setup is designed to keep a detailed record of mergers, which are treated as perfect mergers, using the existing \textit{REBOUND} reb\_collision\_resolve mechanism. We have modified the \textit{REBOUND} source code to keep track of the relative compositions of merged particles (utilizing the ’additional properties’ built-in formalism), which we then use in order to calculate a more realistic physical collision radius for the \textit{REBOUND} particles. Additionally, by the same formalism, we also keep track of the material origin, i.e., if it came from the original impactor or target. 

We modify the code to remove particles that, through close encounters, obtain hyperbolic trajectories during the simulation. This assumption can also be somewhat improved, removing bound particles whose apocenters lie outside $\sim$40\% the planet’s Hill radius \citep{GrishinEtAl-2017}, however, we decide to completely ignore the potential influence of the host star since we have found the number of removed particles to be negligible in the N-body simulations performed in this study. Additionally, particles that enter into the Roche limit are also removed from the simulation. While studies of more energetic moon formation impacts typically use the value for the fluid Roche limit (e.g, \cite{MalamudEtAl-2018}), our current simulations with material strength indicate that for fragments up to and including moonlet-size objects, tidal disruption is inhibited well below the fluid Roche limit. Internally strong fragments can resist tidal disruption down to a fraction of that distance, as anticipated by the rigid Roche limit \citep{Davidsson-1999}, whose value is slightly larger than half of that of the fluid Roche limit. We therefore use the rigid Roche limit when accounting for particle removal.

We run each N-body simulation for a maximum integration time of 350 years. Based on \cite{CitronEtAl-2018}, tidal migration (which is not modeled with our current \textit{REBOUND} code) starts to become important on this time scale, and should not be neglected. If the starting configuration is a field of debris (following a highly disruptive collision), we find that 350 years are more than enough for reaching nearly full coagulation of the debris.

We also note as a caveat that \textit{REBOUND} does not account for the planet's oblateness, which, depending on the timescale of the problem, could be important. Differential precession among the debris can in principle increase their misalignment and in turn the impact velocities of re-collisions. Based on the characteristic time scales \citep{Slotten-2017}, however, we find that typical time between re-collisions (Section \ref{SS:NbodyInterpretation}) is at most comparable with the precession cycle period, and thus significant divergence in inclinations is not expected. While adding the planet's J$_2$ as a free parameter could in principle excite slightly larger collision velocities, the added complexity is probably not significant enough. Yet, this must be kept in mind as a current limitation of our N-body code which treats the planet strictly as a point mass. The resulting impact velocities presented in Section \ref{S:Results} may thus be viewed as a lower limit, but capture the characteristic values.

As mentioned above, occasionally highly disruptive collisions yield a large field of small debris, and we track the coagulation of moonlets from this disk of particles. In most cases, the SPH collision outcomes involve two, mostly intact moonlets, and often also some eroded particles. The amount of eroded debris sensitively depends on the impact velocity and angle. In these cases, we are mainly interested in the dynamical evolution of the two major surviving moonlets inside the debris field. Soft encounters could lead to the infall/ejection of particles, but physical collisions are more important since they alter the mass, mixing, and orbital parameters of the major moonlets. Our \textit{REBOUND} code modifications are designed to keep a record of all physical collisions in the simulations. Since our focus is on the primary two moonlets, whenever a re-collision between them occurs, we hand over their position/velocity coordinates and updated masses back to \textit{miluphcuda}, to perform a subsequent hydrodynamic simulation, as described in Section \ref{SS:Hydro}. During this transfer, we ignore other n-body particles that are not in the two major moonlets since their mass at this stage is typically negligible (after having been accreted already onto the primary moonlets or Earth).

\section{Results}\label{S:Results}
In the following, we categorize the result of each sequence of simulations by four possible outcomes that replace the previous assumption of a perfect merger, as follows:\\

\noindent\textbf{merger} - one moonlet emerged, which perfectly merged the masses of the two original moonlets.\\

\noindent\textbf{growth} - one moonlet emerged, which grew in mass compared to the original most-massive moonlet.\\

\noindent\textbf{erosion} - one moonlet emerged, which lost significant mass relative to the original most massive moonlet.\\


\noindent\textbf{restart (growth/erosion/null)} - Either two moonlets emerged, or one moonlet accompanied by additional debris fragments of non-negligible mass. The most massive of the original moonlets either gained mass (growth), lost mass (erosion), or stayed the same at the level of 1\% (null). Their long-term tidal evolution must be restarted.

\subsection{Synopsis of evolutionary outcomes}\label{SS:Synopsis}
The outcomes of the collisions specified in Table \ref{tab:Cases} are shown in Tables \ref{tab:Merger}-\ref{tab:RestartNull}. Each scenario is tracked in detail and in multiple stages, first by performing a hydrodynamic collision simulation and then by performing as many N-body / hydrodynamic simulations as required, in iterative steps. We continue the process for as many stages as necessary until one of the aforementioned outcomes is achieved. We find that the longest outcomes are resolved in merely 6 stages, three iterations of hydrodynamic simulations and three iterations of N-body simulations. Most outcomes (more than half) are resolved in just 2-3 stages. For each outcome listed above, we provide a separate table: merger (Table \ref{tab:Merger}), growth (Table \ref{tab:Growth}), erosion (Table \ref{tab:Erosion}), restart-growth (Table \ref{tab:RestartGrowth}), restart-erosion (Table \ref{tab:RestartErosion}) and restart-null (Table \ref{tab:RestartNull}).

\begin{table*}[h!]
	\begin{tabular}{|l|c|c|c|c|c|c|c|c|c|c|c|c|c|c|c|}
		\hline
	    & $m_1$ ($M_{\rm L}$) & $a_1$ ($R_\oplus$) & $e_1$ & $P_1$ (h) & $f_{\rm 1<-1}$ & $f_{\rm 1<-2}$ & $m_2$ ($M_{\rm L}$) & $a_2$ ($R_\oplus$) & $e_2$ & $P_2$ (h) & $f_{\rm 2<-1}$ & $f_{\rm 2<-2}$ & $d$ ($R_\oplus$) & $v$ (m/s) & $\theta$ ($^\circ$) \\ \hline
		
		\textbf{21u\_23} & 0.1899 & 19.43 & 0.38 & & 1 & 0 & 0.4712 & -44.99 & 1.54 & & 0 & 1 & 26.6 & 1570 & 7.52 \\
        1st~Hydro & 0.6611 & 132.5 & 0.81 & 15.36 & 0.29 & 0.71 & & & & & & & & & \\ \hline	

        \textbf{21f\_148} & 0.1797 & 17.34 & 0.38 & & 1 & 0 & 0.3098 & 114.59 & 0.84 & & 0 & 1 & 20.1 & 1474 & 23.5 \\
        1st~Hydro & 0.49 & 28.5 & 0.31 & 15.2 & 0.37 & 0.63 & & & & & & & & & \\ \hline

        \textbf{20f\_129} & 0.3249 & 8.931 & 0.14 & & 1 & 0 & 0.2471 & 17.67 & 0.79 & & 0 & 1 & 7.92 & 2895 & 31.9 \\
        1st~Hydro & 0.3215 & 8 & 0.22 & 12.7 & 0.91 & 0.09 & 0.223 & 9.72 & 0.58 & 12.32 & 0.1 & 0.9 & & & \\
        1st~NBody & 0.3226 & 8.04 & 0.22 & & 0.91 & 0.09 & 0.224 & 9.7 & 0.58 & & 0.09 & 0.91 & 8.8 & 1270 & 38 \\
        2nd~Hydro & 0.547 & 8.16 & 0.38 & 4.26 & 0.59 & 0.41 & & & & & & & & & \\ \hline

        \textbf{21m\_53} & 0.1401 & 22.99 & 0.2 & & 1 & 0 & 0.3546 & 19.71 & 0.54 & & 0 & 1 & 19.8 & 1344 & 35.9 \\
        1st~Hydro & 0.4952 & 18.24 & 0.37 & 4.65 & 0.28 & 0.72 & & & & & & & & & \\ \hline

        \textbf{20n\_91} & 0.3527 & 5.5171 & 0.0345 & & 1 & 0 & 0.1467 & 9.84 & 0.42 & & 0 & 1 & 5.71 & 2920 & 48.1 \\
        1st~Hydro & 0.3533 & 5.58 & 0.03 & 16.1 & 0.98 & 0.02 & 0.1387 & 6.58 & 0.19 & 37.1 & 0.02 & 0.98 & & & \\ 
        1st~NBody & 0.3533 & 5.58 & 0.03 & & 0.98 & 0.02 & 0.1387 & 6.58 & 0.19 & & 0.02 & 0.98 & 5.73 & 2200 & 21.7 \\
        2nd~Hydro & 0.492 & 5.34 & 0.08 & 6.83 & 0.72 & 0.28 & & & & & & & & & \\ \hline

        \textbf{20p\_17} & 0.111 & 41.4 & 0.55 & & 1 & 0 & 0.1169 & 14.4 & 0.37 & & 0 & 1 & 19 & 1125 & 55.1 \\
        1st~Hydro & 0.2279 & 19 & 0.01 & 3.69 & 0.49 & 0.51 & & & & & & & & & \\ \hline

        \textbf{21c\_201} & 0.264 & 21 & 0.42 & & 1 & 0 & 0.3155 & 9.4 & 0.61 & & 0 & 1 & 12.2 & 1847 & 69.6 \\
        1st~Hydro & 0.264 & 11.9 & 0.32 & 29.4 & 1 & 0 & 0.3155 & 10.5 & 0.33 & 35.6 & 0 & 1 & & & \\ 
        1st~NBody & 0.264 & 11.9 & 0.32 & & 1 & 0 & 0.3155 & 10.6 & 0.33 & & 0 & 1 & 11.7 & 944 & 67.8 \\
        2nd~Hydro & 0.5795 & 10.7 & 0.33 & 4.62 & 0.46 & 0.53 & & & & & & & & & \\ \hline

        \textbf{21\_32} & 0.2662 & 11.9 & 0.2 & & 1 & 0 & 0.3269 & 48.9 & 0.75 & & 0 & 1 & 12.3 & 3087 & 79.2 \\
        1st~Hydro & 0.2661 & 9.42 & 0.37 & 201 & 1 & 0 & 0.3269 & 44.7 & 0.73 & 216 & 0 & 1 & & & \\ 
        1st~NBody & 0.2661 & 9.48 & 0.38 & & 1 & 0 & 0.3269 & 45.3 & 0.73 & & 0 & 1 & 12.2 & 2684 & 84.6 \\
        2nd~Hydro & 0.2647 & 10.9 & 0.19 & 29.2 & 1 & 0 & 0.3266 & 17.8 & 0.32 & 32.5 & 0 & 1 & & & \\
        2nd~NBody & 0.2647 & 11 & 0.19 & & 1 & 0 & 0.3266 & 17.9 & 0.32 & & 0 & 1 & 12.3 & 2021 & 71.9 \\
        3rd~Hydro & 0.5914 & 11.3 & 0.14 & 3.92 & 0.45 & 0.55 & & & & & & & & & \\
  
		\hline		
	\end{tabular}

    \caption{List of scenarios leading to \textbf{merger}. Similar to Table \ref{tab:Cases}, with six additional columns, provided separately per moonlet $i$: the mass fraction from original moonlet 1 ($f_{i \rm <-1}$), the mass fraction from original moonlet 2 ($f_{i \rm <-2}$) and the rotation period ($P$). When only a single moonlet emerges, it is shown for $i$=1.}
	\label{tab:Merger}
\end{table*}

\begin{table*}[h!]
	\begin{tabular}{|l|c|c|c|c|c|c|c|c|c|c|c|c|c|c|c|}
		\hline
	    & $m_1$ ($M_{\rm L}$) & $a_1$ ($R_\oplus$) & $e_1$ & $P_1$ (h) & $f_{\rm 1<-1}$ & $f_{\rm 1<-2}$ & $m_2$ ($M_{\rm L}$) & $a_2$ ($R_\oplus$) & $e_2$ & $P_2$ (h) & $f_{\rm 2<-1}$ & $f_{\rm 2<-2}$ & $d$ ($R_\oplus$) & $v$ (m/s) & $\theta$ ($^\circ$) \\ \hline

        \textbf{22\_174} & 0.1523 & 30.4 & 0.43 & & 1 & 0 & 0.7809 & 13.5 & 0.59 & & 0 & 1 & 18.4 & 3387 & 71.4 \\
        1st~Hydro & 0.1521 & 14.1 & 0.33 & 31.5 & 1 & 0 & 0.7809 & 13.1 & 0.66 & 192 & 0 & 1 & & & \\ 
        1st~NBody & 0.1521 & 14.1 & 0.33 & & 1 & 0 & 0.7809 & 13.1 & 0.66 & & 0 & 1 & 18.6 & 2751 & 65.3\\
        2nd~Hydro & 0.0001$^{\rm a}$ & 9.86 & 0.81 & 2.1 & 0.08 & 0.92 & 0.8097 & 11.5 & 0.81 & 11.2 & 0.04 & 0.96 & & & \\

		\hline		
	\end{tabular}

    \caption{List of scenarios leading to \textbf{growth}. Column description as in Table \ref{tab:Merger}.
    \newline $^{\rm a}$ The smaller of the two original moonlets transferred some mass to the larger moonlet and fell onto the Earth, leaving negligible bound debris.}
	\label{tab:Growth}
\end{table*}

\begin{table*}[h!]
	\begin{tabular}{|l|c|c|c|c|c|c|c|c|c|c|c|c|c|c|c|}
		\hline
	    & $m_1$ ($M_{\rm L}$) & $a_1$ ($R_\oplus$) & $e_1$ & $P_1$ (h) & $f_{\rm 1<-1}$ & $f_{\rm 1<-2}$ & $m_2$ ($M_{\rm L}$) & $a_2$ ($R_\oplus$) & $e_2$ & $P_2$ (h) & $f_{\rm 2<-1}$ & $f_{\rm 2<-2}$ & $d$ ($R_\oplus$) & $v$ (m/s) & $\theta$ ($^\circ$) \\ \hline

        \textbf{21b\_59} & 0.2754 & 23.53 & 0.71 & & 1 & 0 & 0.3987 & 5.43 & 0.24 & & 0 & 1 & 6.75 & 5211 & 5.17 \\
        1st~Hydro & 0.0678 & 4.65 & 0.59 & 5.58 & 0.46 & 0.54 & 0.0619 & 8.96 & 0.68 & 7.06 & 0.29 & 0.71 & & & \\ 
        1st~NBody & 0.0087$^{\rm a}$ & 13 & 0.76 & & 0.49 & 0.51 & 0.3362 & 5.33 & 0.48 & & 0.39 & 0.61 & & & \\

		\hline		
	\end{tabular}

    \caption{List of scenarios leading to \textbf{erosion}. Column description as in Table \ref{tab:Merger}.
    \newline $^{\rm a}$ A catastrophic collision resulted in a field of debris, of which only one large moonlet emerged, and the second-largest fragment is of insignificant mass.}
	\label{tab:Erosion}
\end{table*}

\begin{table*}[h!]
	\begin{tabular}{|l|c|c|c|c|c|c|c|c|c|c|c|c|c|c|c|}
		\hline
	    & $m_1$ ($M_{\rm L}$) & $a_1$ ($R_\oplus$) & $e_1$ & $P_1$ (h) & $f_{\rm 1<-1}$ & $f_{\rm 1<-2}$ & $m_2$ ($M_{\rm L}$) & $a_2$ ($R_\oplus$) & $e_2$ & $P_2$ (h) & $f_{\rm 2<-1}$ & $f_{\rm 2<-2}$ & $d$ ($R_\oplus$) & $v$ (m/s) & $\theta$ ($^\circ$) \\ \hline
  
		\textbf{20c\_35} & 0.1204 & 8.335 & 0.36 & & 1 & 0 & 0.2039 & 6.11 & 0.29 & & 0 & 1 & 5.36 & 3481 & 18.7 \\
        1st~Hydro & 0.0245 & 4.708 & 0.27 & 3.73 & 0.8 & 0.2 & 0.1961 & 5.52 & 0.14 & 5.6 & 0.21 & 0.79 & & & \\ 
        1st~NBody & 0.0529 & 4.35 & 0.2 & & 0.71 & 0.29 & 0.2277 & 5.57 & 0.14 & & 0.26 & 0.74 & & & \\ \hline

        \textbf{20f\_101} & 0.4342 & 6.69 & 0.19 & & 1 & 0 & 0.1927 & 6.23 & 0.5 & & 0 & 1 & 6.56 & 2017 & 40.6 \\
        1st~Hydro & 0.4552 & 6.6 & 0.11 & 9.23 & 0.93 & 0.07 & 0.1714 & 4.62 & 0.18 & 9.43 & 0.06 & 0.94 & & & \\ 
        1st~NBody & 0.4554 & 6.58 & 0.1 & & 0.93 & 0.07 & 0.1716 & 4.69 & 0.17 & & 0.06 & 0.94 & & & \\ \hline

        \textbf{20\_166} & 0.2265 & 5.64 & 0.23 & & 1 & 0 & 0.1988 & 14.88 & 0.59 & & 0 & 1 & 6.43 & 4361 & 57.3 \\
        1st~Hydro & 0.2234 & 5.21 & 0.27 & 11.6 & 0.99 & 0.01 & 0.1951 & 12.9 & 0.53 & 34.8 & 0.01 & 0.99 & & & \\ 
        1st~NBody & 0.225 & 5.21 & 0.27 & & 0.99 & 0.01 & 0.1951 & 12.9 & 0.53 & & 0.01 & 0.99 & 6.44 & 3937 & 20.6 \\
        2nd~Hydro & 0.1513 & 4.89 & 0.32 & 9.67 & 0.52 & 0.48 & 0.0847 & 4.37 & 0.45 & 3.3 & 0.88 & 0.12 & & & \\
        2nd~NBody & 0.2725 & 4.73 & 0.35 & & 0.63 & 0.37 & 0.0729 & 5.66 & 0.19 & & 0.23 & 0.77 & & & \\ \hline

        \textbf{21\_215} & 0.2208 & 22.7 & 0.62 & & 1 & 0 & 0.328 & 7.5 & 0.16 & & 0 & 1 & 8.61 & 3828 & 65.4 \\
        1st~Hydro & 0.2199 & 20.5 & 0.58 & 34.5 & 1 & 0 & 0.3278 & 6.7 & 0.29 & 38.9 & 0 & 1 & & & \\ 
        1st~NBody & 0.2199 & 20.6 & 0.58 & & 1 & 0 & 0.3279 & 6.73 & 0.29 & & 0 & 1 & 8.72 & 3392 & 55 \\
        2nd~Hydro & 0.193 & 9.69 & 0.09 & 14.9 & 0.96 & 0.04 & 0.3229 & 6.7 & 0.33 & 8.91 & 0.05 & 0.95 & & & \\
        2nd~NBody & 0.1936 & 9.71 & 0.09 & & 0.96 & 0.04 & 0.3522 & 6.44 & 0.34 & & 0.05 & 0.95 & 8.75 & 2086 & 81 \\
        3rd~Hydro & 0.1895 & 8.39 & 0.12 & 17.7 & 0.96 & 0.04 & 0.3552 & 6.15 & 0.41 & 19.8 & 0.05 & 0.95 & & & \\
        3rd~NBody & 0.1936 & 8.47 & 0.12 & & 0.96 & 0.04 & 0.3559 & 6.16 & 0.41 & & 0.05 & 0.95 & & & \\
  
		\hline		
	\end{tabular}

    \caption{List of scenarios leading to \textbf{restart (growth)}. Column description as in Table \ref{tab:Merger}.}
	\label{tab:RestartGrowth}
\end{table*}

\begin{table*}
	\begin{tabular}{|l|c|c|c|c|c|c|c|c|c|c|c|c|c|c|c|}
		\hline
	    & $m_1$ ($M_{\rm L}$) & $a_1$ ($R_\oplus$) & $e_1$ & $P_1$ (h) & $f_{\rm 1<-1}$ & $f_{\rm 1<-2}$ & $m_2$ ($M_{\rm L}$) & $a_2$ ($R_\oplus$) & $e_2$ & $P_2$ (h) & $f_{\rm 2<-1}$ & $f_{\rm 2<-2}$ & $d$ ($R_\oplus$) & $v$ (m/s) & $\theta$ ($^\circ$) \\ \hline
  
        \textbf{20f\_143} & 0.2282 & 3.58 & 0.02 & & 0 & 1 & 0.1525 & 102.8 & 0.97 & & 1 & 0 & 3.65 & 5215 & 12.84 \\
        1st~Hydro & 0.0067 & 3.42 & 0.17 & 11.5 & 0.77 & 0.23 & 0.0051 & 3.4 & 0.05 & 3.49 & 0.66 & 0.34 & & & \\ 
        1st~NBody & 0.1609 & 3.26 & 0.15 & & 0.6 & 0.4 & 0.144 & 3.86 & 0.04 & & 0.7 & 0.3 & & & \\

		\hline		
	\end{tabular}

    \caption{List of scenarios leading to \textbf{restart (erosion)}. Column description as in Table \ref{tab:Merger}.}
	\label{tab:RestartErosion}
\end{table*}

\begin{table*}
	\begin{tabular}{|l|c|c|c|c|c|c|c|c|c|c|c|c|c|c|c|}
		\hline
	    & $m_1$ ($M_{\rm L}$) & $a_1$ ($R_\oplus$) & $e_1$ & $P_1$ (h) & $f_{\rm 1<-1}$ & $f_{\rm 1<-2}$ & $m_2$ ($M_{\rm L}$) & $a_2$ ($R_\oplus$) & $e_2$ & $P_2$ (h) & $f_{\rm 2<-1}$ & $f_{\rm 2<-2}$ & $d$ ($R_\oplus$) & $v$ (m/s) & $\theta$ ($^\circ$) \\ \hline
  
		\textbf{20n\_53} & 0.3034 & 6.7 & 0.17 & & 1 & 0 & 0.2622 & 21.87 & 0.71 & & 0 & 1 & 7.88 & 4223 & 12.35 \\
        1st~Hydro & 0.2631 & 5.4 & 0.44 & 5.05 & 0.67 & 0.33 & 0.0915 & 6.75 & 0.4 & 4.95 & 0.32 & 0.68 & & & \\ 
        1st~NBody & 0.297 & 5.33 & 0.43 & & 0.71 & 0.29 & 0.149 & 6.82 & 0.34 & & 0.26 & 0.74 & & & \\ \hline 

        \textbf{20f\_118} & 0.2137 & 10.04 & 0.69 & & 1 & 0 & 0.1454 & 7.8 & 0.18 & & 0 & 1 & 8.88 & 2769 & 27.5 \\
        1st~Hydro & 0.2136 & 7.68 & 0.59 & 9.54 & 0.88 & 0.12 & 0.122 & 6.75 & 0.4 & 4.95 & 0.12 & 0.88 & & & \\ 
        1st~NBody & 0.219 & 7.66 & 0.59 & & 0.87 & 0.13 & 0.127 & 6.72 & 0.39 & & 0.14 & 0.86 & & & \\ \hline

        \textbf{20\_2} & 0.4405 & 6 & 0.37 & & 1 & 0 & 0.2747 & 44.96 & 0.83 & & 0 & 1 & 7.75 & 4572 & 28.2 \\
        1st~Hydro & 0.3729 & 5.1 & 0.5 & 16.1 & 0.91 & 0.09 & 0.122 & 9.08 & 0.23 & 13.5 & 0.11 & 0.89 & & & \\ 
        1st~NBody & 0.443 & 4.95 & 0.5 & & 0.83 & 0.17 & 0.197 & 7.92 & 0.17 & & 0.19 & 0.81 & & & \\ \hline

        \textbf{20c\_46} & 0.3013 & 6.67 & 0.25 & & 1 & 0 & 0.1763 & 16.9 & 0.74 & & 0 & 1 & 7.66 & 4077 & 84.1 \\
        1st~Hydro & 0.3013 & 6.74 & 0.25 & 997 & 1 & 0 & 0.1763 & 11.8 & 0.68 & 429 & 0 & 1 & & & \\ 
        1st~NBody & 0.3013 & 6.76 & 0.25 & & 1 & 0 & 0.1763 & 11.8 & 0.68 & & 0 & 1 & 5.41 & 5290 & 75.9 \\
        2nd~Hydro & 0.2975 & 6.48 & 0.26 & 57.9 & 1 & 0 & 0.1718 & 9.2 & 0.6 & 41.3 & 0 & 1 & & & \\
        2nd~NBody & 0.2989 & 6.49 & 0.26 & & 1 & 0 & 0.1736 & 9.1 & 0.6 & & 0 & 1 & 5.19 & 5098 & 73.1 \\
        3rd~Hydro & 0.2906 & 5 & 0.24 & 18.2 & 1 & 0 & 0.1676 & 8.94 & 0.56 & 18.9 & 0 & 1 & & & \\
        3rd~NBody & 0.2954 & 4.98 & 0.23 & & 0.99 & 0.01 & 0.1695 & 8.76 & 0.55 & & 0.01 & 0.99 & & & \\ \hline

        \textbf{20e\_19} & 0.2288 & 13.6 & 0.47 & & 1 & 0 & 0.3628 & 54.8 & 0.67 & & 0 & 1 & 18.7 & 1454 & 87 \\
        1st~Hydro & 0.2287 & 17.7 & 0.22 & 176 & 1 & 0 & 0.3628 & 25.1 & 0.22 & 266 & 0 & 1 & & & \\ 
        1st~NBody & 0.2287 & 17.4 & 0.24 & & 1 & 0 & 0.3628 & 25.1 & 0.23 & & 0 & 1 & & & \\
        
		\hline		
	\end{tabular}

    \caption{List of scenarios leading to \textbf{restart (null)}. Column description as in Table \ref{tab:Merger}.}
	\label{tab:RestartNull}
\end{table*}

\subsection{Moonlet collisions versus coagulation}\label{SS:NbodyInterpretation}
In Tables \ref{tab:Merger}-\ref{tab:RestartNull}, whenever $d$, $v$, and $\theta$ are displayed following an N-body calculation (impact distance, velocity and angle respectively), it indicates that two moons have collided. We did not specify the time that elapsed between the start of the N-body simulation and the collision, due to space limitations in Tables \ref{tab:Merger}-\ref{tab:RestartNull}, but we found that when collisions did occur, the time varied between several days to years. Whenever $d$, $v$, and $\theta$ are not displayed, it is either because the simulation tracked the coagulation of moonlets from a field of small debris (e.g., scenarios 21b\_59 and 20f\_143, where small collisions in mostly co-rotating debris are assumed to be perfect mergers), or else there was simply no collision between major moonlets for the entire integration time of 350 years (e.g., scenarios 20c\_35, 20f\_101, 20f\_118, 20e\_19 and others). There are also some in-between cases such as scenarios 20\_166, 20n\_53, and 20\_2, in which one or two of the moonlets are partially rather than fully eroded.

Figure \ref{fig:20_166} below shows a special case as an example (scenario 20\_166) where the collision outcome is achieved after 4 stages, and each stage belongs to a different category. First, a hydrodynamic simulation resolves the collision between two similar-sized moonlets at a relatively grazing impact angle and high impact velocity. This graze\&run type impact does not erode the moonlets significantly. The subsequent N-body simulation indicates that the two moonlets collide again after 219 days. The 2nd collision between the two moonlets has nearly the same velocity, but the impact angle is almost head-on, which results in a partial erosion of the moonlets, whose mass splits between three large fragments and a lot of additional debris. The last N-body simulation calculates the coagulation of this debris field into one major moonlet, much more massive than the original most-massive moonlet, in addition to a second minor moonlet. During this stage, most collisions indeed involve small debris. However, there is also one notable collision between two of the three large fragments. We find that this collision occurs at an extremely low velocity of merely 522 m $\times$ s$^{-1}$. This velocity is a few times lower than typical velocities that, according to our analysis below, lead to perfect mergers (approximately 1/3 of the mutual escape velocity). Thus, an additional hydrodynamic simulation to resolve this collision is not required. 

\begin{figure*}
    \begin{tabular}[b]{c}
        \includegraphics[width=0.3\textwidth]{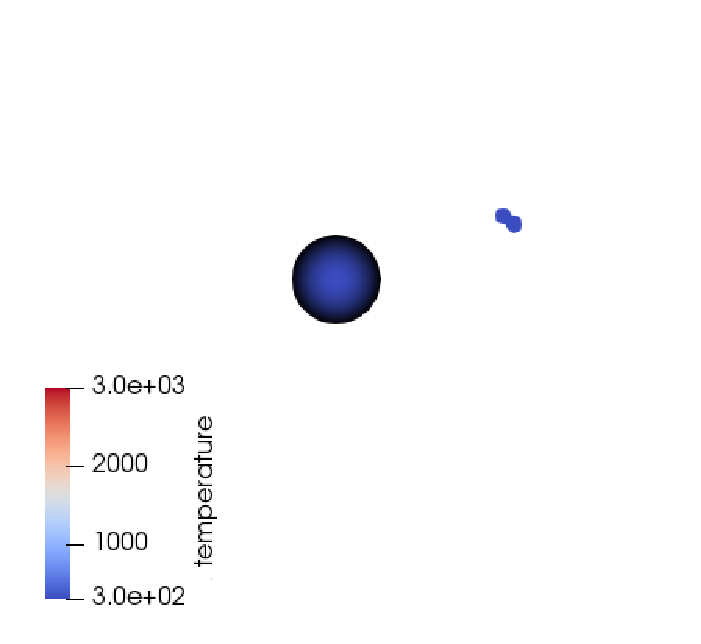}\label{fig:1st_hydro_first}\\
        \small (a) incipient hydrodynamic collision
    \end{tabular}
    \begin{tabular}[b]{c}
	   \includegraphics[width=0.3\textwidth]{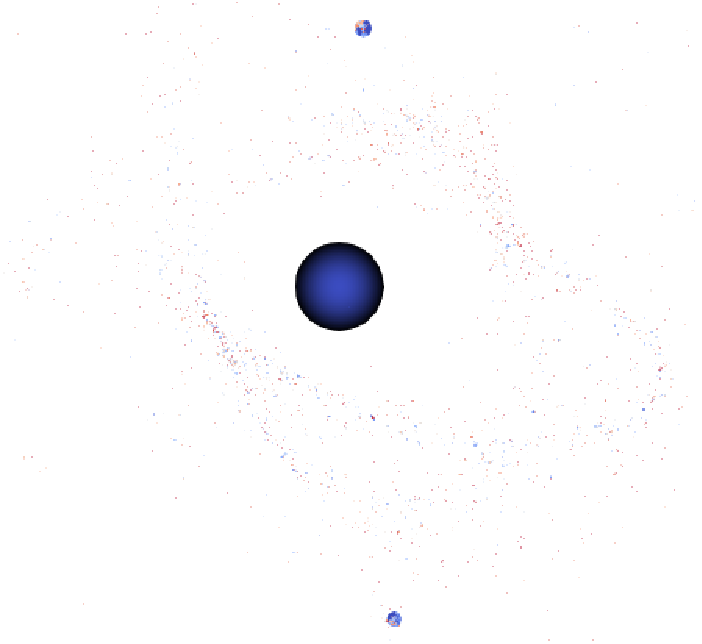}\label{fig:1st_hydro_last}\\
        \small (b) after 28 hours, graze\&run
    \end{tabular}
    \begin{tabular}[b]{c}
	   \includegraphics[width=0.3\textwidth]{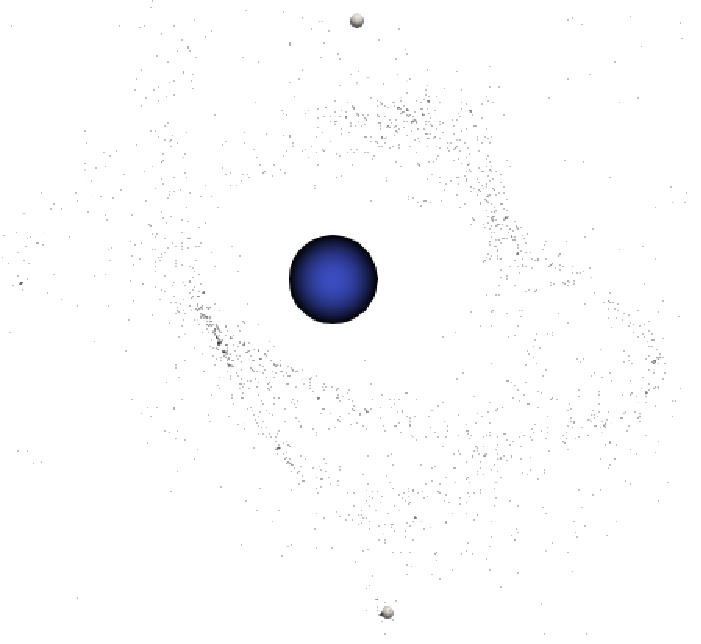}\label{fig:1st_rebound_first}\\
        \small (c) hand over to N-body
    \end{tabular}
    \begin{tabular}[b]{c}
	   \includegraphics[width=0.3\textwidth]{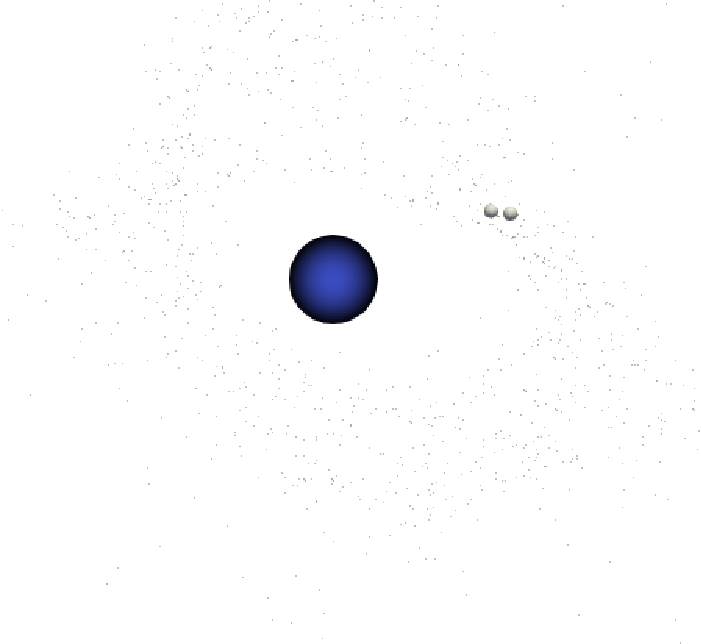}\label{fig:1st_rebound_last}\\
        \small (d) after 219 days, re-collision
    \end{tabular}
    \begin{tabular}[b]{c}
        \includegraphics[width=0.3\textwidth]{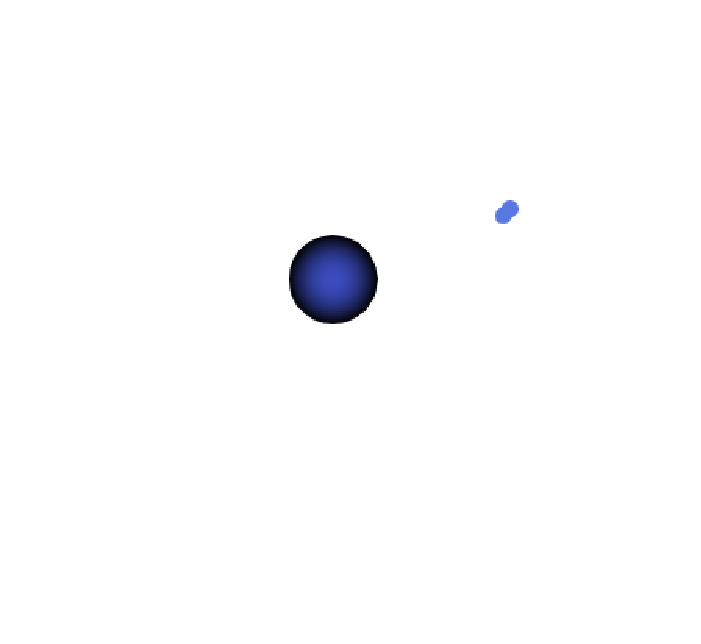}\label{fig:2nd_hydro_first}\\
        \small (e) 2nd hydrodynamic collision
    \end{tabular}
    \begin{tabular}[b]{c}
	   \includegraphics[width=0.3\textwidth]{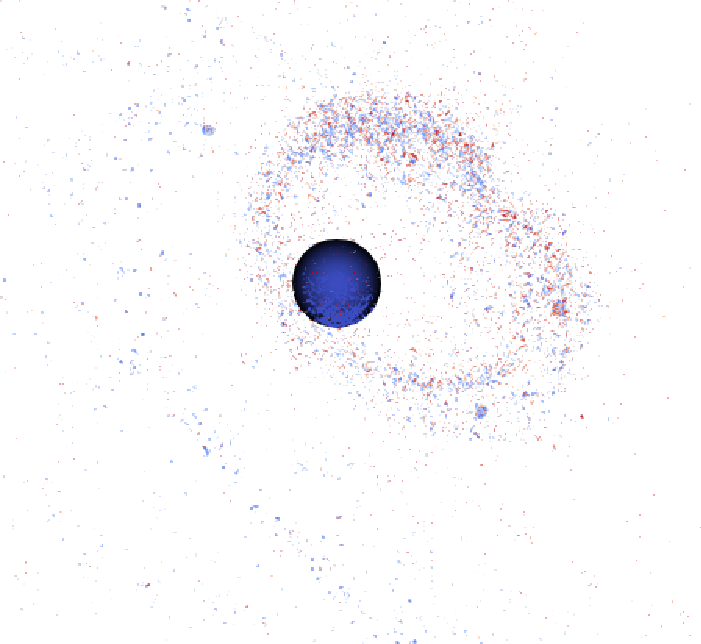}\label{fig:2nd_hydro_last}\\
        \small (f) after 18 hours, partial disruption
    \end{tabular}
    \begin{tabular}[b]{c}
	   \includegraphics[width=0.3\textwidth]{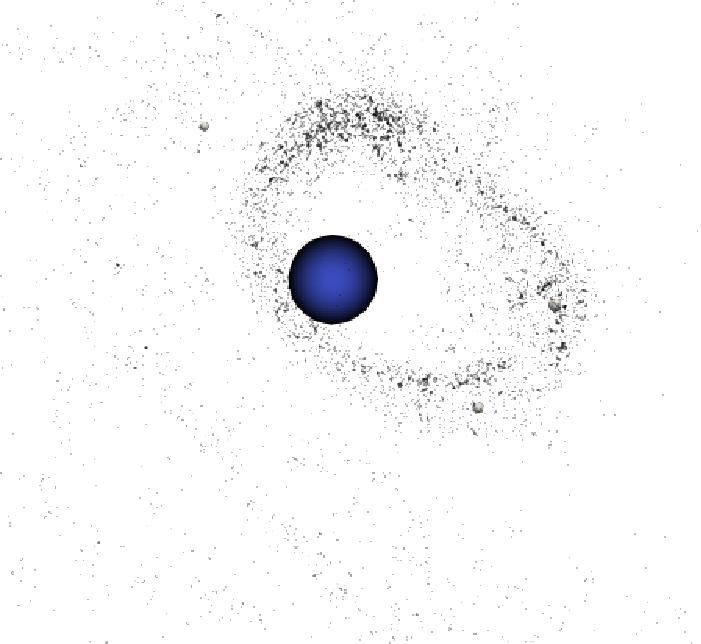}\label{fig:2nd_rebound_first}\\
        \small (g) 2nd hand over to N-body
    \end{tabular}
    \begin{tabular}[b]{c}
	   \includegraphics[width=0.3\textwidth]{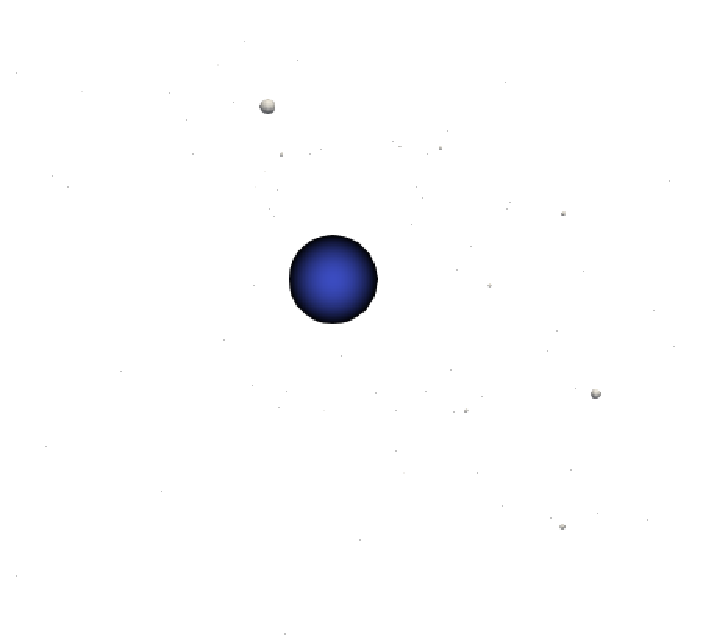}\label{fig:2nd_rebound_last}\\
        \small (h) after 350 years, coagulation
    \end{tabular}
    
    \caption{The 4-stage collision evolution of scenario 20\_166. Each two panels show the respective start and end state of each stage: (a)-(b) 1st hydrodynamic; (c)-(d) 1st N-body; (e)-(f) 2nd hydrodynamic; (g)-(h) 2nd N-body. In the hydrodynamic simulations, the Earth is a single gravitating SPH particle. For visual clarity, it is rendered as a sphere and scaled to the Earth's size. In the N-body simulations, all particles are rendered as spheres and scaled to their respective physical size, which the code is modified to inform. The color scheme for the SPH particles denotes temperature, as shown in the color bar of panel (a).}
    \label{fig:20_166}
\end{figure*}

\subsection{Rotation periods}\label{SS:RotationPeriod}
The rotation of each fragment is calculated as a post-processing step to the hydrodynamic simulations. The process includes calculating the angular momentum vector $L$ of each fragment (the latter being a physically connected clump of SPH particles, found using a friends-of-friends algorithm). Then the moment of inertia tensor $I$ about the center of mass of each fragment is calculated, and in turn, the angular velocity vector $\omega$ is calculated as the dot product, such that $\omega=I^{-1} \cdot L$. The rotation period is then $2\pi / |\omega|$. In order to perform the calculation, each fragment must be resolved by multiple particles, whereas in the N-body calculation, each fragment consists of a single \textit{REBOUND} particle. Hence, the rotational properties in Tables \ref{tab:Merger}-\ref{tab:RestartNull} are only applicable to the analysis of the SPH simulation outcomes, and must be reset during each N-body calculation. While certain modifications might in principle be made in the future to analytically approximate the rotational evolution during mergers or soft encounters inside the N-body code, this remains a task for subsequent studies. In the current study, the periods obtained in each hydrodynamic simulation are always under the assumption of zero rotation in the SPH initial conditions, and thus a limitation of the current study is that rotational outcomes are detached from previous evolutionary stages.

\begin{figure}[h!]
    \centering
    \includegraphics[width=\linewidth]{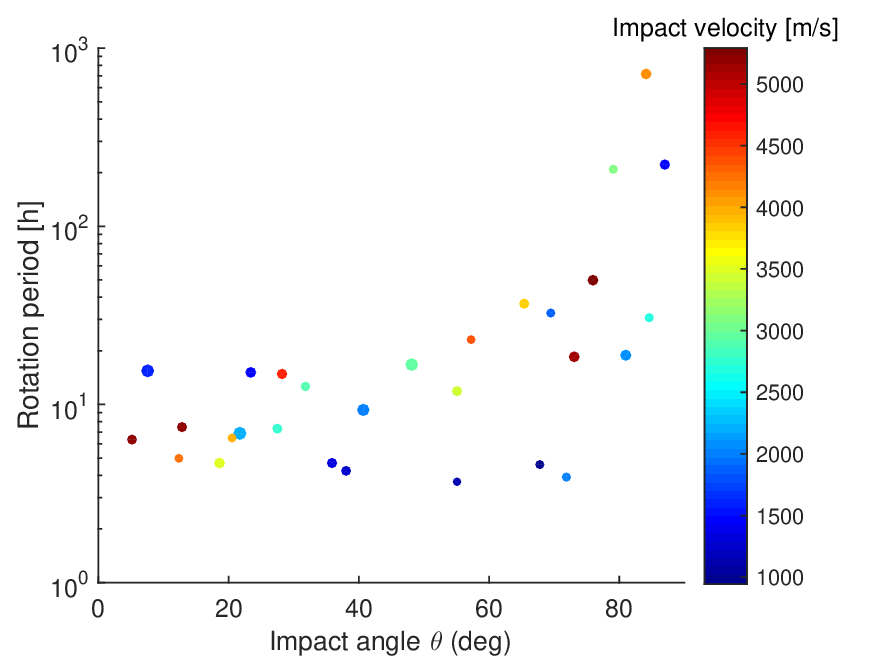}
    \caption{The rotation period of moonlets emerging in hydrodynamic impacts from Table \ref{tab:Cases}, as a function of the impact angle of the original colliding moonlets. The data points are color-coded to the impact velocity (color bar) and size-coded to the mass ratio of the original moonlets (smallest/largest circles representing a 1:1/2.5:1 mass ratio, respectively).}
    \label{fig:PeriodVsAngle}
\end{figure}

Given these limitations, we plot the rotational data obtained for the hydrodynamic simulations in Tables \ref{tab:Merger}-\ref{tab:RestartNull}. Figure \ref{fig:PeriodVsAngle} shows the rotation period of moonlets in our simulations as a function of the impact angle. Whenever there is more than one ensuing moonlet in a given scenario, we take the average of the rotation periods of the two listed moonlets (the mass ratios of both original and ensuing moonlets are always comparable or vary by at most a factor of 2.5). We omit one case in which both the original and ensuing moonlets have a larger mass ratio (22\_174). Each data point is represented by a circle, where each circle is color-coded to the impact velocity and size-coded to the mass ratio of the original moonlets in the impact.

Below an impact angle of around 60$^\circ$, we find that the rotation period of moonlets is generally on the order of 10 h, with some scatter around this value. Above this impact angle, both the scatter and the maximum rotation period tend to increase. These results cannot be understood by a simple analytical model, such as for a perfect plastic merger, since the kinetic energy in the impact is transferred to multiple fragments. The mass ratio of the colliding moonlets or their impact velocity does not appear to correlate with the rotation period in any obvious way. We also check for a correlation (not shown) with the Earth distance during collision, and find none.

\subsection{Ejected and Earth infall material fractions}\label{SS:EjectedInfall}
The relative fractions of bound/ejected(unbound)/Earth infall material are calculated during each stage, but are not listed in Tables  \ref{tab:Merger}-\ref{tab:RestartNull} due to space limitations. However, we note that typically the debris is in bound orbit, and the fraction of ejected or Earth infall material is negligible. There are a number of exceptions to this rule. Among the hydrodynamic simulations, the following scenarios have these outlier fractions (ejection/infall, respectively): 21b\_59 1st (0.05/0.13), 20f\_143 1st (0.08/0.04), 20n\_53 1st (0.04/0.01), 20\_2 1st (0.02/0.015) and 20\_166 2nd (0.02/0.02). Among the N-body simulations, the ejected fraction is always either negligible or zero. However, the Earth infall fraction is not: 21b\_59 (0.32), 20f\_143 1st (0.03), 20n\_53 1st (0.025), 20\_2 1st (0.01) and 20\_166 2nd (0.03), noting that in the N-body simulations, the collision cross-section is the rigid Roche limit (as discussed in Section \ref{SS:NBody}) and not the radius of the Earth (because the SPH simulations can resolve tidal disruptions explicitly while the N-body simulations cannot). Thus, since the rigid Roche radius is about 50\% larger than the Earth's physical radius, the N-body fractions are enhanced. These outlier cases all have in common that the collision velocity is extremely high ($>$4200 m/s) and the angle is not too grazing. In all other cases, assuming that all debris is bound was found to be justified. 

Another unique case is 22\_174 2nd, where following an initial collision, the subsequent N-body integration showed that the two moonlets re-collide after about a month. The second collision takes place near the apocenter of the less-massive moonlet, and orbital energy is extracted from both moonlets such that the less-massive moonlet infalls almost completely intact onto the Earth. The other moonlet survives tidal disruption when passing between the rigid and fluid Roche limit. The resilience to tidal disruption due to internal cohesion is discussed next.

\subsection{Strength versus pure hydrodynamics}\label{SS:StrengthVersusHydro}
In addition to the model realizations that include internal strength (as described in Section \ref{SS:Hydro}), for each scenario in Table \ref{tab:Cases} we also run a purely hydrodynamic simulation. One of our goals is to determine the importance of strength in the simulation outcomes. This includes the effects on the properties of the debris or moonlets that form in collisions, in addition to the tidal effects by the planet.

\begin{figure}[h!]
    \centering
    \includegraphics[width=\linewidth]{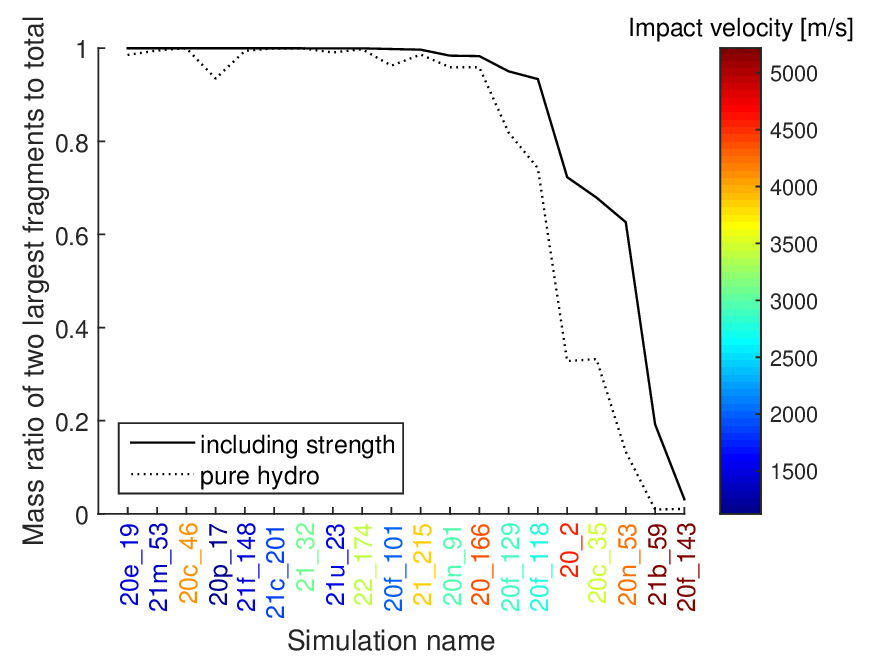}
    \caption{The mass fraction of the two largest fragments (or one fragment in the case of a perfect merger) to the total mass, compared between simulations with internal strength (solid) versus pure hydrodynamics (dotted). The simulation label names are color coded to the impact velocity (colorbar), indicating a general correlation with fragmentation.}
    \label{fig:StrengthVsPure}
\end{figure}

Our initial expectation is that fragmentation would be enhanced in simulations that do not include strength. Figure \ref{fig:StrengthVsPure} shows that indeed when internal strength is neglected, the ratio of the two most massive fragments in the debris (or trivially, one massive fragment in the case of a perfect merger) to that of the initial colliding moonlets, is lower. Cross-referencing these results with Table \ref{tab:Cases}, shows that simulations at lower velocities tend to yield little to no fragmentation. In these cases, internal strength only matters in terms of the final shape/rotation properties of the ensuing perfectly merged moonlets. The difference in fragmentation is only apparent for impact velocities in excess of about 2000 m$\times$s$^{-1}$ (scenario 20f\_101). From this point on, strength-less simulations generate a larger number of (smaller) debris, and this trend increases with the impact velocity. There are three outlier cases to the left of scenario 20f\_101, where weak fragmentation was indicated, despite the fact that the initial impact velocity was higher than 2000 m$\times$s$^{-1}$. These scenarios are outliers for a similar reason. The moonlets impact at extremely grazing impact angle, resulting in impacts which are not very disruptive despite their relatively large impact velocities. 

In Section \ref{SS:NBody}, we stated that when internally strong fragments enter the Roche limit, they should fully tidally disrupt only when considering the rigid Roche limit $R_{\rm Roche\_r}$, whereas the fluid Roche limit $R_{\rm Roche\_f}$, commonly used in various studies, is more judicious in the case of pure hydrodynamic simulations or when considering larger scales for which gravity is completely dominant. This decision is supported in the literature \citep{Davidsson-1999}. However, here we also test its validity by comparing tidal disruption cases with and without strength. Having looked in our simulation suite, we found that scenario 20\_2 (Table \ref{tab:RestartNull}) is highly appropriate for this validation test. The 20\_2 impact simulation showed that the most massive moon $m_1$ has a pericenter distance of 1.64 $\times$ 10$^7$ m, which we find to be equivalent to 0.88 $R_{\rm Roche\_f}$. One would expect in this case to see significant tidal stripping in the fluid case, and no tidal deformation in the solid case, since the rigid Roche limit $R_{\rm Roche\_r}$ is slightly over 0.5 $R_{\rm Roche\_f}$. 

\begin{figure*}[h!]
    \begin{tabular}[b]{c}
        \includegraphics[width=0.49\textwidth]{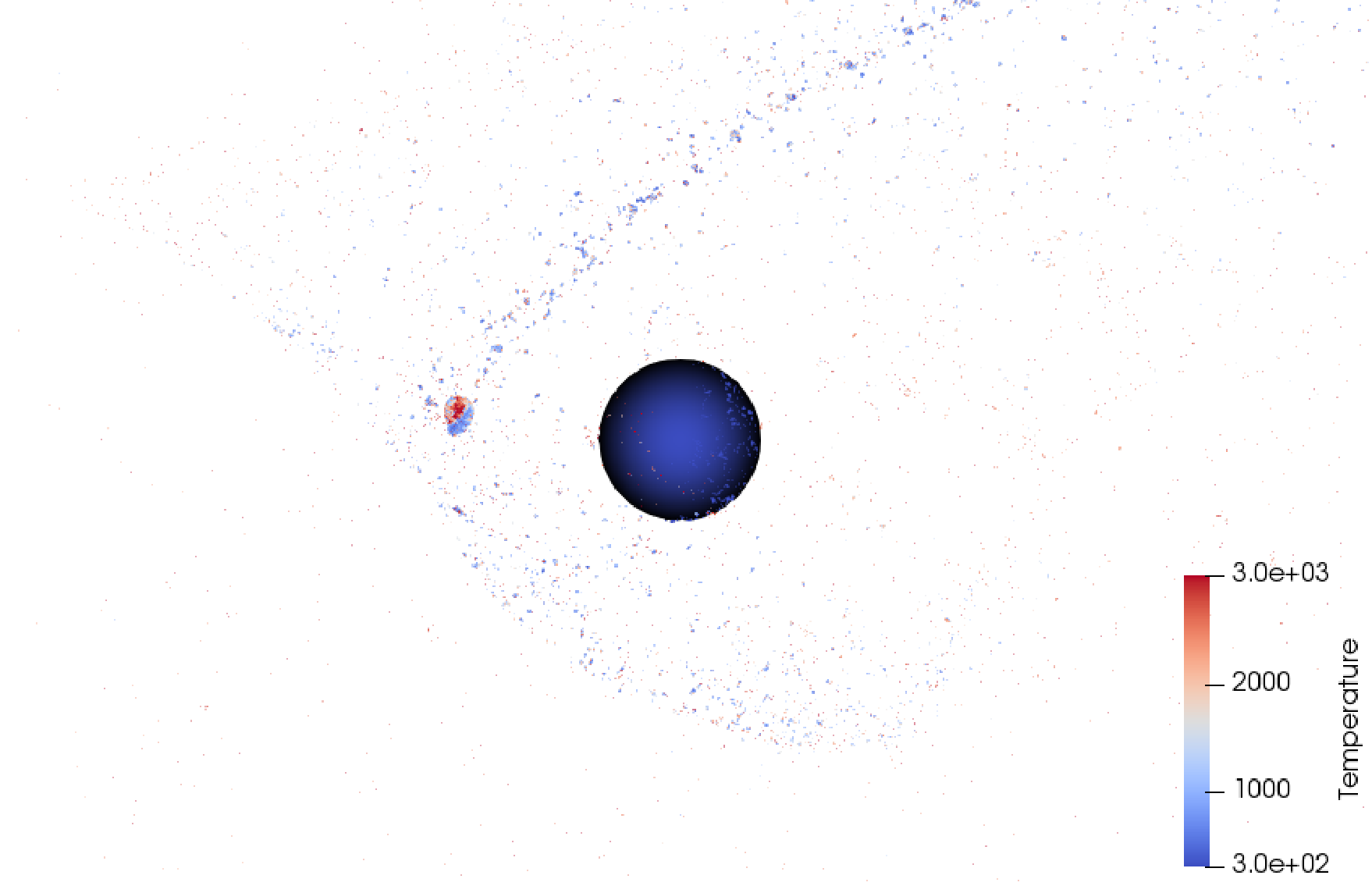}\label{fig:20_2_solid_no_tidal_elongation_6_94h}\\
        \small (a) including strength, 6.94 h, no tidal elongation
    \end{tabular}
    \begin{tabular}[b]{c}
	   \includegraphics[width=0.49\textwidth]{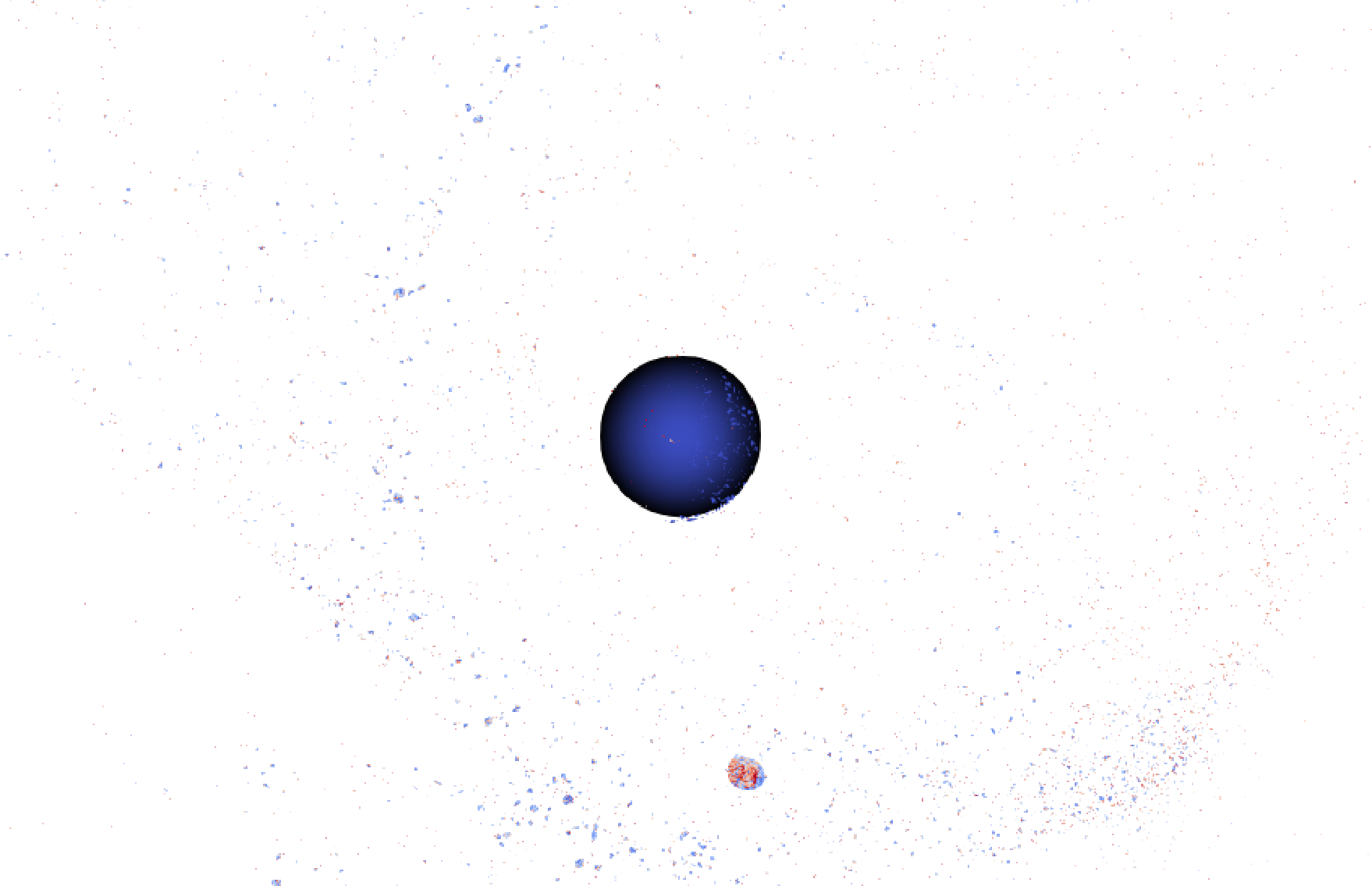}
        \label{fig:20_2_solid_no_tidal_stripping_10_42h}\\
        \small (b) including strength, 10.42 h, no tidal stripping, $\Delta q=0$
    \end{tabular}
    \begin{tabular}[b]{c}
	   \includegraphics[width=0.49\textwidth]{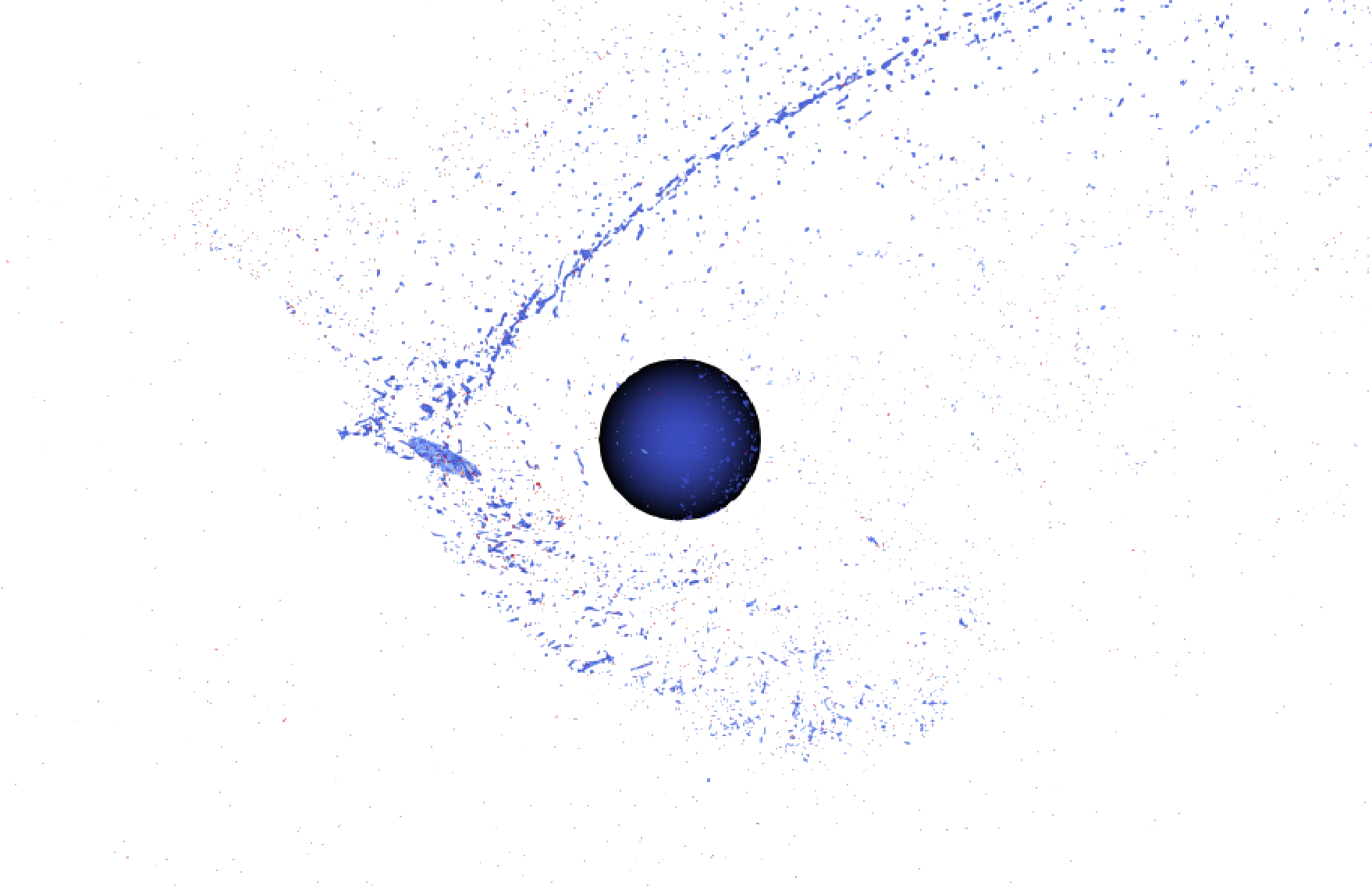}\label{fig:20_2_solid_tidal_elongation_6_94h}\\
        \small (c) pure hydrodynamics, 6.94 h, 1st tidal elongation
    \end{tabular}
    \begin{tabular}[b]{c}
	   \includegraphics[width=0.49\textwidth]{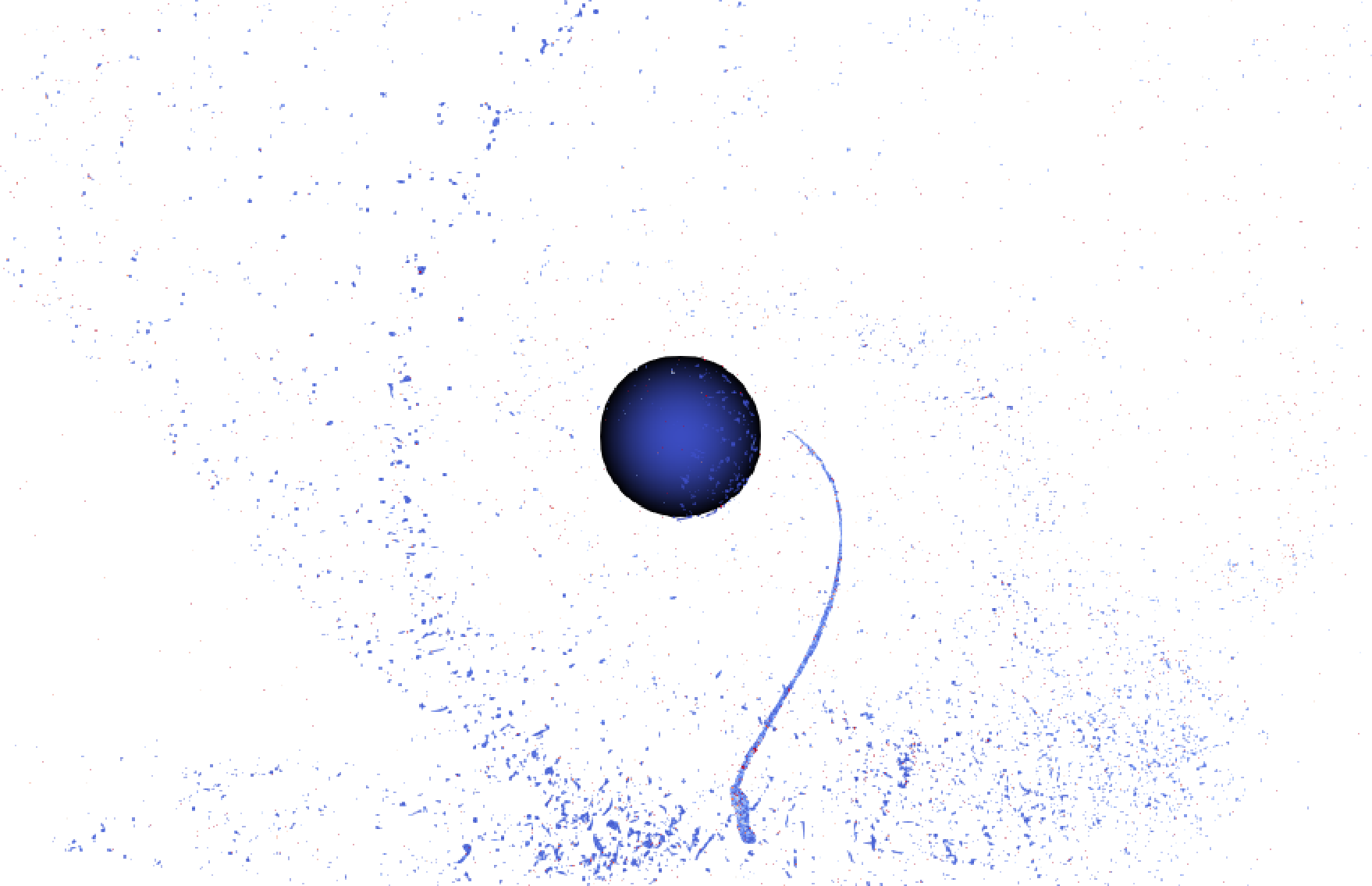}\label{fig:20_2_solid_tidal_stripping_10_42h}\\
        \small (d) pure hydrodynamics, 10.42 h, 1st tidal stripping, $\Delta q \uparrow 25\%$
    \end{tabular}
    \begin{tabular}[b]{c}
        \includegraphics[width=0.49\textwidth]{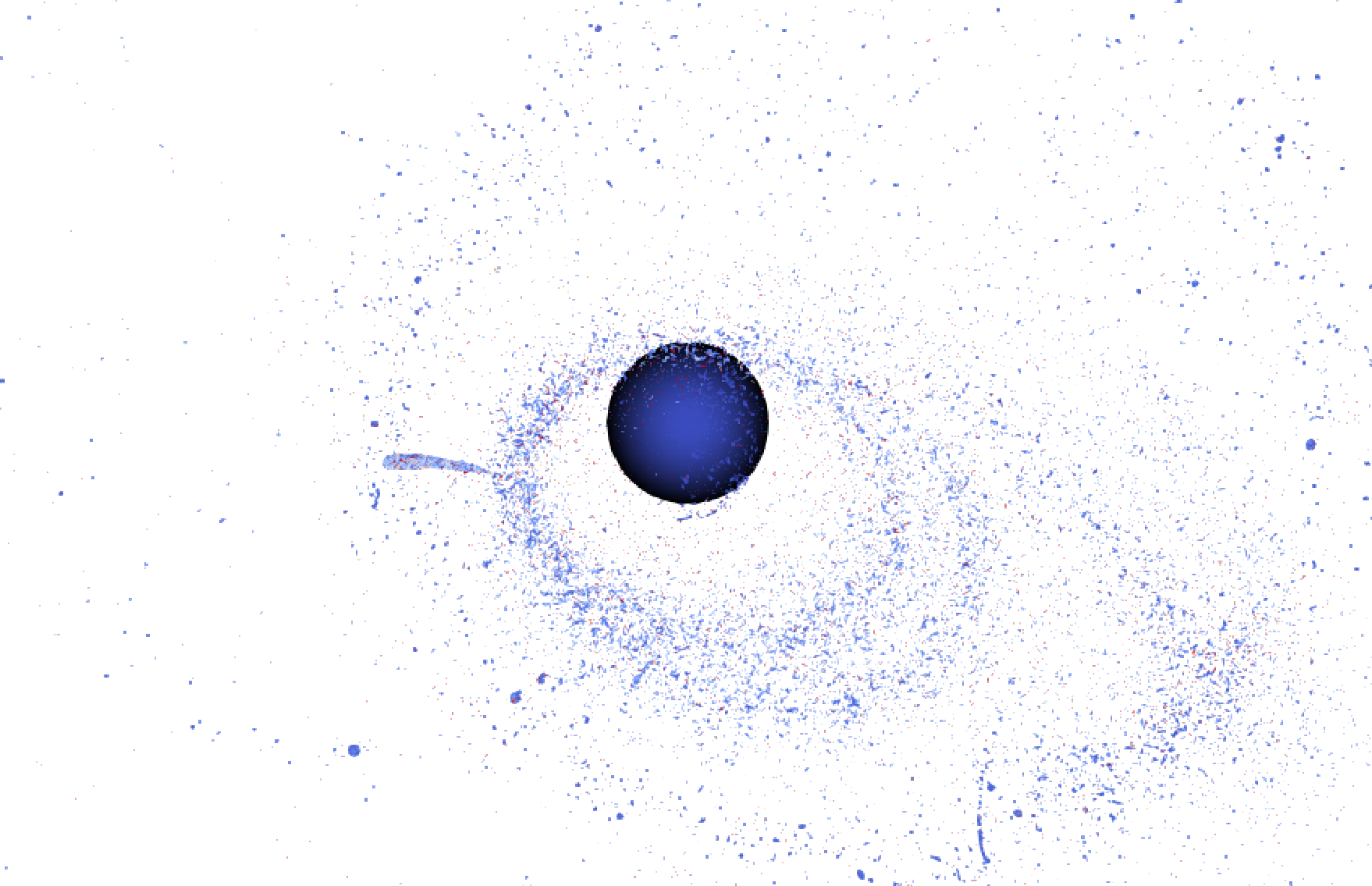}\label{fig:20_2_solid_tidal_elongation_30_56h}\\
        \small (e) pure hydrodynamics, 30.56 h, 2nd tidal elongation
    \end{tabular}
    \begin{tabular}[b]{c}
	   \includegraphics[width=0.49\textwidth]{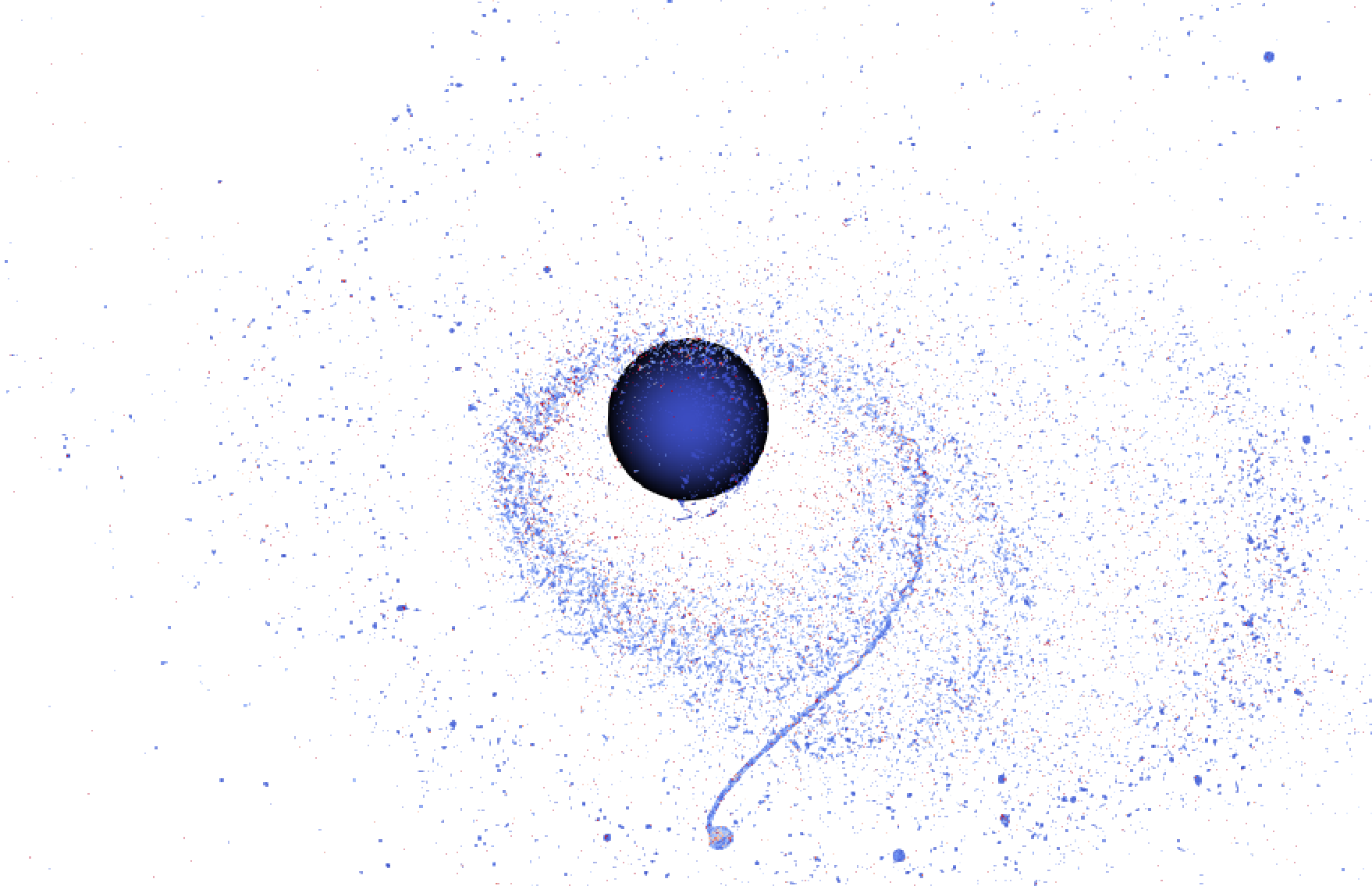}\label{fig:20_2_solid_tidal_strippingn_34_03h}\\
        \small (f) pure hydrodynamics, 34.03 h, 2st tidal stripping, $\Delta q \uparrow 10\%$
    \end{tabular}
    
    \caption{Close Earth approach of the massive moonlet ($m_1$) from scenario 20\_2 in Table \ref{tab:RestartNull}. The Earth is a single gravitating SPH particle, rendered as a sphere and scaled to the Earth's size. The orbital pericenter $q$ is left of the Earth. The 1st ($q=0.88 R_{\rm Roche\_f}$ -- (a),(c)) and 2nd ($q=1.12 R_{\rm Roche\_f}$ -- (e)) close approaches occur 6.94 and 30.56 h after the initial collision (not shown), respectively. Panels: (a) 1st close approach including internal strength; (b) $q$ remains unaffected several hours later; (c) 1st close approach with pure hydrodynamics indicating tidal elongation; (d) tidal stripping is observed several hours later, after which the pericenter $q$ increases by 25\%; (e) 2nd close approach with pure hydrodynamics indicating tidal elongation; (f) tidal stripping is observed a second time, after which the pericenter $q$ increases by additional 10\%. The color scheme for the SPH particles denotes temperature, as shown in the color bar of panel (a).}
    \label{fig:TidalDisruption}
\end{figure*}

Figure \ref{fig:TidalDisruption} indeed confirms this expectation. The top two panels show no change to either the moonlet's elongation or its orbit. The middle and bottom panels show the 1st and 2nd pericenter pass, respectively, in the purely hydrodynamic case. Here in both cases, tidal stripping is observed, and the pericenter increases to 2.06 $\times$ 10$^7$ m and then 2.25 $\times$ 10$^7$ m, respectively, after each close approach (even tough the second close approach occurred outside the fluid Roche limit at 1.1 $R_{\rm Roche\_f}$). Clearly, a model realization without internal strength leads to dramatically different results. In the fluid case, both the orbit and mass of the moonlet were altered considerably. The vicinity of the planet was filled with a debris field from the mass stripping (in two pericenter passes the moonlet has lost more than 68\% of its initial mass). In contrast, no change occurred in the default simulation with internal strength. We note that alterations in the strength model's parameters could make a difference. Here we only considered one model realization, and more comparisons with different model realizations would be beneficial in future work.

We also tested the close approach outcome of an internally strong moonlet that formed in scenario 21b\_59 (not shown), since it had a much closer pericenter distance of merely 1.24 $\times$ 10$^7$ m, still larger than yet now in close proximity to $R_{\rm Roche\_r}$. We found that with the current model parameters there was indeed some partial degree of tidal elongation and stripping, enough to erode 14\% of the moonlet's mass. We conclude based on both scenarios, that close to the Roche limit (either rigid or fluid), partial tidal disruption is to be expected -- for significant mass stripping, the analytic Roche expressions reasonably accurately depict the outcome.

When comparing the fractions of ejected and Earth infall material in strength versus pure hydrodynamic simulations, we find that they increase considerably in the latter. The fractions we listed in Section \ref{SS:EjectedInfall} for our default runs with internal strength (ejected/Earth infall, respectively) have increased as follows: 21b\_59 1st (0.05/0.13 $\rightarrow$ 0.09/0.15), 20f\_143 1st (0.08/0.04 $\rightarrow$ 0.12/0.05), 20n\_53 1st (0.04/0.01 $\rightarrow$ 0.07/0.045), 20\_2 1st (0.02/0.015 $\rightarrow$ 0.045/0.02) and 20\_166 2nd (0.02/0.02 $\rightarrow$ 0.045/0.045). In this case, too, we conclude that internal strength correlates with increased consolidation.

\subsection{Fragmentation and tidal forces}\label{SS:FragmantationEarthDistance}
Whereas in the previous section, we discussed the effect of internal strength on fragmentation, here we emphasize the role of tidal forces, and in turn the distance to the planet during collision. Our expectation is to obtain increased fragmentation with rising tidal forces, and in turn when the collision takes place closer to the Earth. 

A suitable pair of cases for performing an examination of the tidal environment's effects are scenarios 21b\_59 versus 20f\_143. Both of these collisions are characterized by almost the same mass ratio between the two moonlets ($\sim 2/3$), a low impact angle, and an identical impact velocity. Only the impact distance from the Earth (6.75 versus 3.65 $R_\oplus$, respectively) is markedly different in these simulations. Even though the impact angle in 21b\_59 is slightly smaller and thus closer to a head-on collision, the two ensuing largest fragments in this collision are an order of magnitude more massive than those in 20f\_143. The total mass in the original moonlets of 21b\_59 is only twice as large as in 20f\_143, so there is a factor five enhancement to fragmentation in the latter simulation, consequent of the fact that less energy is required to disrupt the moonlets closer to the Earth.

\section{Discussion}\label{S:Discussion}
This work considers multiple collisions among moonlets during the late stages of accretion of the Earth. Unlike previous studies, we employ self-consistent initial conditions from prior numerical modeling, rather than performing a parameter study that uses a grid of ad-hock initial impact geometries. Using this approach, we have shown that the naive assumption of a perfect merger, often made in N-body calculations of moonlet dynamics, and which has also been made in past dynamical studies (e.g., \citep{HaghighipourRaymond-2007,FischerCiesla-2014}), including for the moon's formation via multiple impacts \citep{CitronEtAl-2018}, is in many cases incorrect. As one might expect a priori, impact velocities below (or slightly above) the mutual escape velocity of the colliding moonlets lead to perfect mergers. At any higher velocities, we observe that impact outcomes are more challenging to predict with accuracy, identifying several collision categories.

At the high end of the impact velocity range, when the velocity is over about 2.5 times the mutual escape velocity, the collision is predictably disruptive (unless the impact angle is extremely grazing). The outcome, however, is hard to predict. Scenarios 21b\_59 and 20f\_143 both result in debris disks, however, the former coagulates into a single moonlet, while the latter into two moonlets. In both outcomes, the most massive moonlet is eroded compared to the original most massive moonlet, and therefore these outcomes may be classified as erosive. In 20f\_143, since two moonlets have emerged, and since they did not collide in 350 years (the timescale before tidal migration becomes important) the subsequent longer-term dynamical evolution must be considered, which exceeds the scope of the current work.

Our simulations have confirmed that perfect mergers occur for velocities up to 1-1.2 times the mutual escape velocity and that the most disruptive collisions occurred at velocities around 2.7 times the mutual escape velocity. These results agree with past estimations (e.g., \cite{Asphaug-2010,LeinhardtStewart-2012}) when a central planet is not considered. As we have seen, the effects of the Earth are more subtle and are manifested through the accretion of mass from the debris (Section \ref{SS:EjectedInfall}), enhancing fragmentation and sometimes enabling tidal disruption on Roche limit interior orbits (Section \ref{SS:FragmantationEarthDistance}). The challenging part is to predict outcomes in the large range of intermediate impact velocities. This is a known problem in simulations involving collisions in free space \citep{Chambers-2013,ClementEtAl-2019}, and it is complicated even further near the central gravitational potential of a planet.

For intermediate impact velocities, we find that there are several possibilities. Sometimes a perfect merger still occurs, albeit following several iterations. Scenarios 20f\_129, 20n\_91, 21c\_201 and 21\_32 (Table \ref{tab:Merger}), in which the impact velocity is below about twice the mutual escape velocity, have collided several times following hit\&run sequences but remained on orbits that still crossed paths. Sequential collisions were shown to damp both semi-major axes and impact velocities, eventually leading to the outcome of perfect mergers. However, many scenarios are characterized by a series of collisions that do not eventually lead to a merger, but at most to some minor mass transfer. Among these cases (Tables \ref{tab:RestartGrowth} and \ref{tab:RestartNull}), the last such collision results in stable moonlet orbits that avoid mutual collisions for hundreds of years. As mentioned in Section \ref{SS:NBody}, we find that tidal migration must characteristically be taken into account at longer time scales, and therefore these simulations must be restarted, which exceeds the scope of this work. 

Perhaps the most significant detail shown by our simulations, in virtually all 'restart' cases, is that the semi-major axes of the moonlets tend to decrease during and after collisions. Likewise, material mixing is often also exhibited. We have not employed any post-processing algorithms to look at surface mixing following hydrodynamic simulations. Nor do we expect this analysis to remain precisely the same over the long term since various processes such as convection, radiogenic heating, relaxation, melting, and resurfacing are expected to modify the surface characteristics on geological timescales post-impact. Nevertheless, the bulk material fractions in the final merged or re-assembled moonlets indicate that mixing between restarted simulations is important, and is likely to imprint on the surface mixing as well. Note that orbital damping and mixing occur mainly in the hydrodynamical collision simulations. The potential of the N-body simulations to damp the orbital parameters further, enhance impactor-projectile mixing and transfer mass, increases as $v$ increases, since then more debris is generated.

Fortunately, we find here that tracking the full course of the collisional evolution is feasible within typically a few iterations. This finding is true even in those cases where the impact angle is extremely grazing, although typically this extends the sequence a little bit. This result is encouraging for future numerical computations involving the multiple impact scenario hypothesis, because a future framework can be established where the direct impact outcomes following tidal evolution \citep{CitronEtAl-2018} can be further informed by linking these N-body simulations to hydrodynamic ones, at a relatively reasonable computational cost.

In a previous study by \cite{CitronEtAl-2018}, it was found that in the framework of multiple impacts, each impact had more or less a similar chance of resulting in the survival of only the more massive outer moonlet (which was the more tidally evolved product of past interactions), increase its mass via a merger, or disrupt the system by removing both of the moonlets. Hence, the growth potential, either increasing the mass of the most massive moonlet or keeping it unaltered, was generally found to be about twice that of disrupting the system. The full sequence of growth up to the contemporary mass of the moon was deemed possible.

However, one of our main goals in the current study was to re-evaluate these aforementioned probabilities by revising the past assumption of a perfect merger. We find here that collisions previously assumed as perfect mergers, may in fact (with respect to the most massive moonlet) branch into either perfect mergers, non-perfect but mass-increasing outcomes, null (unaltered mass) outcomes, or erosive outcomes. Only the latter kind, erosive outcomes, negate growth. Nevertheless, even then, the sequence may potentially continue in the direction of growth, since erosion, unlike full disruption of the system, is not a terminal outcome. Our current findings suggest that in 20 simulated cases, only 2 were erosive. Although we are dealing with small statistics, the current study suggests that the sequence of growth is generally maintained, even if perfect mergers branch into several distinct outcomes. 

We have also tested the influence of internal strength in hydrodynamic simulations, here for the first time to our knowledge near a central gravitational potential. We compared these results with identical pure hydrodynamic simulations. We find that including strength is particularly relevant for moonlet-sized objects. We demonstrate that strength inhibits fragmentation, and is particularly important for the more energetic collisions at intermediate and high impact velocity. We find that strength inhibits tidal disruption, or rather that the rigid Roche limit must be considered in the framework of N-body simulations, and not the fluid Roche limit. In light of this, past numerical studies that considered tidal evolution and dynamics of moonlets in N-body simulations, should probably be revised, since the inner radius where satellites were assumed to be destroyed (and accreted to the Earth) due to disruption in N-body multiple moon simulations, was significantly larger (about twice) than the realistic disruption radius that we found here. Hence, a larger fraction of satellites that were considered to be destroyed in previous simulations should have survived and contributed to the build-up of larger moonlets. We find that strength reduces the fraction of ejected and Earth-infall material in collisions. Given the above, we conclude that future research on moon formation (either ours or other terrestrial / extra-terrestrial moons) should not ignore strength in hydrodynamic simulations.

Some additional aspects of moon-moon collisions in the multiple impact hypothesis are left for future follow-up studies. For example, here we did not consider the planet's oblateness, as this is currently a limitation of the N-body code we are using. Further consideration of this complication in the future, might excite slightly higher collision velocities in the N-body stage, and in turn slightly increase fragmentation, mixing and reduction of semi-major axes. Additionally, we did not calculate the melt fraction generated in collisions, and instead chose to focus on the merger outcome, reflecting on both the impact parameters and the modeling methodology. To calculate realistic melt fractions, which are important in the context of the formation of the moon's anorthositic crust \citep{RusselEtAl-2014,SchwingerBreuer-2022}, we require better treatment of the initial thermal state in the hydrodynamic simulations (as shown by \cite{RufuAharonson-2019}, and we expect it to be even more significant with internal strength). The initial state as well as the state between subsequent impacts requires a combination of software modifications that exceed the current scope of the paper, including energy-tracking modifications to our N-body code and magma cooling calculations.

In the framework of the moon's formation via multiple impacts, previous studies have focused separately on either hydrodynamic simulations \citep{RufuEtAl-2017,MalamudEtAl-2018,RufuAharonson-2019} or N-body simulations \citep{CitronEtAl-2018}. Each of these studies targeted a particular aspect of the problem, either the incipient formation of moonlet-forming disks, the tidal migration of moonlets, the infall of moonlets onto Earth, or their mutual collisions. Here we have employed an approach that combines these two different numerical techniques. While we are still focusing on one aspect - mutual moonlet collisions - the tools developed here can now be advanced further by extending them to other aspects of the problem, and/or by employing additional numerical methods such as geophysical evolution calculations of the moonlets themselves, in between subsequent interactions. These directions will be addressed in future studies.

\section{Acknowledgments}\label{S:Acknowledgments}
We wish to thank Robert Citron for providing us with a list of impact parameters used in a predecessor study. These were used for generating self-consistent initial conditions. UM and HBP acknowledge support from the Israeli Ministry of Science and Technology MOST-space grant and the Minerva Center for Life Under Extreme Conditions.

\newpage
\bibliographystyle{apj}
\bibliography{bibfile} 

\begin{thebibliography}{41}
\expandafter\ifx\csname natexlab\endcsname\relax\def\natexlab#1{#1}\fi

\bibitem[{{Agnor} {et~al.}(1999){Agnor}, {Canup}, \& {Levison}}]{Agn+99}
{Agnor}, C.~B., {Canup}, R.~M., \& {Levison}, H.~F. 1999, \icarus, 142, 219

\bibitem[{{Asphaug}(2010)}]{Asphaug-2010}
{Asphaug}, E. 2010, Chemie der Erde / Geochemistry, 70, 199

\bibitem[{{Asphaug}(2014)}]{Asp+14}
---. 2014, Annual Review of Earth and Planetary Sciences, 42, 551

\bibitem[{{Benz} \& {Asphaug}(1999)}]{BenzAsphaug-1999}
{Benz}, W. \& {Asphaug}, E. 1999, \icarus, 142, 5

\bibitem[{{Bolmont} {et~al.}(2015){Bolmont}, {Raymond}, {Leconte}, {Hersant}, \& {Correia}}]{BolmontEtAl-2015}
{Bolmont}, E., {Raymond}, S.~N., {Leconte}, J., {Hersant}, F., \& {Correia}, A. C.~M. 2015, \aap, 583, A116

\bibitem[{{Briaud} {et~al.}(2023){Briaud}, {Ganino}, {Fienga}, {M{\'e}min}, \& {Rambaux}}]{BriaudEtAl-2023}
{Briaud}, A., {Ganino}, C., {Fienga}, A., {M{\'e}min}, A., \& {Rambaux}, N. 2023, \nat, 617, 743

\bibitem[{{Burger} {et~al.}(2018){Burger}, {Maindl}, \& {Sch{\"a}fer}}]{BurgerEtAl-2018}
{Burger}, C., {Maindl}, T.~I., \& {Sch{\"a}fer}, C.~M. 2018, Celestial Mechanics and Dynamical Astronomy, 130, 2

\bibitem[{{Canup} {et~al.}(1999){Canup}, {Levison}, \& {Stewart}}]{Can+99}
{Canup}, R.~M., {Levison}, H.~F., \& {Stewart}, G.~R. 1999, \aj, 117, 603

\bibitem[{{Canup} {et~al.}(2023){Canup}, {Righter}, {Dauphas}, {Pahlevan}, {{\'C}uk}, {Lock}, {Stewart}, {Salmon}, {Rufu}, {Nakajima}, \& {Magna}}]{Can+23}
{Canup}, R.~M., {Righter}, K., {Dauphas}, N., {Pahlevan}, K., {{\'C}uk}, M., {Lock}, S.~J., {Stewart}, S.~T., {Salmon}, J., {Rufu}, R., {Nakajima}, M., \& {Magna}, T. 2023, Reviews in Mineralogy and Geochemistry, 89, 53

\bibitem[{{Chambers}(2013)}]{Chambers-2013}
{Chambers}, J.~E. 2013, \icarus, 224, 43

\bibitem[{{Citron} {et~al.}(2014){Citron}, {Aharonson}, {Perets}, \& {Genda}}]{Cit+14}
{Citron}, R.~I., {Aharonson}, O., {Perets}, H., \& {Genda}, H. 2014, in 45th Annual Lunar and Planetary Science Conference, Lunar and Planetary Science Conference, 2085

\bibitem[{{Citron} {et~al.}(2018){Citron}, {Perets}, \& {Aharonson}}]{CitronEtAl-2018}
{Citron}, R.~I., {Perets}, H.~B., \& {Aharonson}, O. 2018, \apj, 862, 5

\bibitem[{{Citron} \& {Stewart}(2022)}]{Cit+22}
{Citron}, R.~I. \& {Stewart}, S.~T. 2022, \psj, 3, 116

\bibitem[{{Clement} {et~al.}(2019){Clement}, {Kaib}, {Raymond}, {Chambers}, \& {Walsh}}]{ClementEtAl-2019}
{Clement}, M.~S., {Kaib}, N.~A., {Raymond}, S.~N., {Chambers}, J.~E., \& {Walsh}, K.~J. 2019, \icarus, 321, 778

\bibitem[{{Collins} {et~al.}(2004){Collins}, {Melosh}, \& {Ivanov}}]{CollinsEtAl-2004}
{Collins}, G.~S., {Melosh}, H.~J., \& {Ivanov}, B.~A. 2004, \maps, 39, 217

\bibitem[{{{\'C}uk} \& {Stewart}(2012)}]{CukStewart-2012}
{{\'C}uk}, M. \& {Stewart}, S.~T. 2012, Science, 338, 1047

\bibitem[{{Davidsson}(1999)}]{Davidsson-1999}
{Davidsson}, B. J.~R. 1999, \icarus, 142, 525

\bibitem[{{Emsenhuber} {et~al.}(2024){Emsenhuber}, {Asphaug}, {Cambioni}, {Gabriel}, {Schwartz}, {Melikyan}, \& {Denton}}]{EmsenhuberEtAl-2024}
{Emsenhuber}, A., {Asphaug}, E., {Cambioni}, S., {Gabriel}, T. S.~J., {Schwartz}, S.~R., {Melikyan}, R.~E., \& {Denton}, C.~A. 2024, \psj, 5, 59

\bibitem[{{Fischer} \& {Ciesla}(2014)}]{FischerCiesla-2014}
{Fischer}, R.~A. \& {Ciesla}, F.~J. 2014, Earth and Planetary Science Letters, 392, 28

\bibitem[{{Genda} {et~al.}(2012){Genda}, {Kokubo}, \& {Ida}}]{GendaEtAl-2012}
{Genda}, H., {Kokubo}, E., \& {Ida}, S. 2012, \apj, 744, 137

\bibitem[{{Grishin} {et~al.}(2017){Grishin}, {Perets}, {Zenati}, \& {Michaely}}]{GrishinEtAl-2017}
{Grishin}, E., {Perets}, H.~B., {Zenati}, Y., \& {Michaely}, E. 2017, \mnras, 466, 276

\bibitem[{{Haghighipour} \& {Raymond}(2007)}]{HaghighipourRaymond-2007}
{Haghighipour}, N. \& {Raymond}, S.~N. 2007, \apj, 666, 436

\bibitem[{{Hosono} {et~al.}(2019){Hosono}, {Karato}, {Makino}, \& {Saitoh}}]{Hos+19}
{Hosono}, N., {Karato}, S.-i., {Makino}, J., \& {Saitoh}, T.~R. 2019, Nature Geoscience, 12, 418

\bibitem[{{Jutzi}(2015)}]{Jutzi-2015}
{Jutzi}, M. 2015, \planss, 107, 3

\bibitem[{{Jutzi} \& {Asphaug}(2011)}]{JutziAsphaug-2011}
{Jutzi}, M. \& {Asphaug}, E. 2011, \nat, 476, 69

\bibitem[{{Leinhardt} \& {Stewart}(2012)}]{LeinhardtStewart-2012}
{Leinhardt}, Z.~M. \& {Stewart}, S.~T. 2012, \apj, 745, 79

\bibitem[{{Lock} {et~al.}(2018){Lock}, {Stewart}, {Petaev}, {Leinhardt}, {Mace}, {Jacobsen}, \& {Cuk}}]{loc+18}
{Lock}, S.~J., {Stewart}, S.~T., {Petaev}, M.~I., {Leinhardt}, Z., {Mace}, M.~T., {Jacobsen}, S.~B., \& {Cuk}, M. 2018, Journal of Geophysical Research (Planets), 123, 910

\bibitem[{{Malamud} {et~al.}(2018){Malamud}, {Perets}, {Sch{\"a}fer}, \& {Burger}}]{MalamudEtAl-2018}
{Malamud}, U., {Perets}, H.~B., {Sch{\"a}fer}, C., \& {Burger}, C. 2018, \mnras, 479, 1711

\bibitem[{{Mastrobuono-Battisti} \& {Perets}(2017)}]{Mas+17}
{Mastrobuono-Battisti}, A. \& {Perets}, H.~B. 2017, \mnras, 469, 3597

\bibitem[{{Mastrobuono-Battisti} {et~al.}(2015){Mastrobuono-Battisti}, {Perets}, \& {Raymond}}]{mas+15}
{Mastrobuono-Battisti}, A., {Perets}, H.~B., \& {Raymond}, S.~N. 2015, \nat, 520, 212

\bibitem[{{Melosh}(2007)}]{Melosh-2007}
{Melosh}, H.~J. 2007, \maps, 42, 2079

\bibitem[{{Ogawa} {et~al.}(2021){Ogawa}, {Nakamura}, {Suzuki}, \& {Hasegawa}}]{OgawaEtAl-2021}
{Ogawa}, R., {Nakamura}, A.~M., {Suzuki}, A.~I., \& {Hasegawa}, S. 2021, \icarus, 362, 114410

\bibitem[{{Ringwood}(1989)}]{Rin89}
{Ringwood}, A.~E. 1989, Earth and Planetary Science Letters, 95, 208

\bibitem[{{Rufu} \& {Aharonson}(2019)}]{RufuAharonson-2019}
{Rufu}, R. \& {Aharonson}, O. 2019, Journal of Geophysical Research (Planets), 124, 1008

\bibitem[{{Rufu} {et~al.}(2017){Rufu}, {Aharonson}, \& {Perets}}]{RufuEtAl-2017}
{Rufu}, R., {Aharonson}, O., \& {Perets}, H.~B. 2017, Nature Geoscience, 10, 89

\bibitem[{{Russell} {et~al.}(2014){Russell}, {Joy}, {Jeffries}, {Consolmagno}, \& {Kearsley}}]{RusselEtAl-2014}
{Russell}, S.~S., {Joy}, K.~H., {Jeffries}, T.~E., {Consolmagno}, G.~J., \& {Kearsley}, A. 2014, Philosophical Transactions of the Royal Society of London Series A, 372A, 20130241

\bibitem[{{Sch{\"a}fer} {et~al.}(2020){Sch{\"a}fer}, {Wandel}, {Burger}, {Maindl}, {Malamud}, {Buruchenko}, {Sfair}, {Audiffren}, {Vavilina}, \& {Winter}}]{SchaferEtAl-2020}
{Sch{\"a}fer}, C.~M., {Wandel}, O.~J., {Burger}, C., {Maindl}, T.~I., {Malamud}, U., {Buruchenko}, S.~K., {Sfair}, R., {Audiffren}, H., {Vavilina}, E., \& {Winter}, P.~M. 2020, Astronomy and Computing, 33, 100410

\bibitem[{{Schwinger} \& {Breuer}(2022)}]{SchwingerBreuer-2022}
{Schwinger}, S. \& {Breuer}, D. 2022, Physics of the Earth and Planetary Interiors, 322, 106831

\bibitem[{{Slotten}(2017)}]{Slotten-2017}
{Slotten}, J.~D. 2017, in 7th European Conference on Space Debris, 131

\bibitem[{{Speith}(2006)}]{Speith-2006}
{Speith}, R. 2006, habilitation Thesis

\bibitem[{{Weber} {et~al.}(2011){Weber}, {Lin}, {Garnero}, {Williams}, \& {Lognonn{\'e}}}]{WeberEtAl-2011}
{Weber}, R.~C., {Lin}, P.-Y., {Garnero}, E.~J., {Williams}, Q., \& {Lognonn{\'e}}, P. 2011, Science, 331, 309

\end{thebibliography}

\appendix

\section{Appendix A: Convergence test}\label{Appendix:A}
In the main text we discussed our choice to use lower resolution SPH simulations for the most disruptive impacts. Our decision was based on performance considerations for both the SPH simulations and the follow-up N-body simulations, which are considerably slowed as a result of increasing the number of debris fragments (and hence the number of N-body particles).

In order to investigate the effect of resolution on SPH outcomes, we ran some of the simulations at multiple resolutions. In Figure we show an example for two of the simulations, 20\_2 and 20f\_143, which were set up at four different resolutions: 50k,100k,250k and 500k SPH particles. These two scenarios are interesting, due to their highly and extremely highly disruptive nature, respectively. Here we show the mass (panel (a)) and rotation period (panel (b)) of the largest fragment in the debris, evolving as a function of time, up to more than 15 hours post-impact, which is a sufficient time for convergence. As explained in the main text, the colliding moonlets are set up as non-rotating. Since the initial rotation period is infinite, we start panel (b) after 3 hours, for visual clarity.

\begin{figure*}[h!]
    \begin{tabular}[b]{c}
        \includegraphics[width=0.507\textwidth]{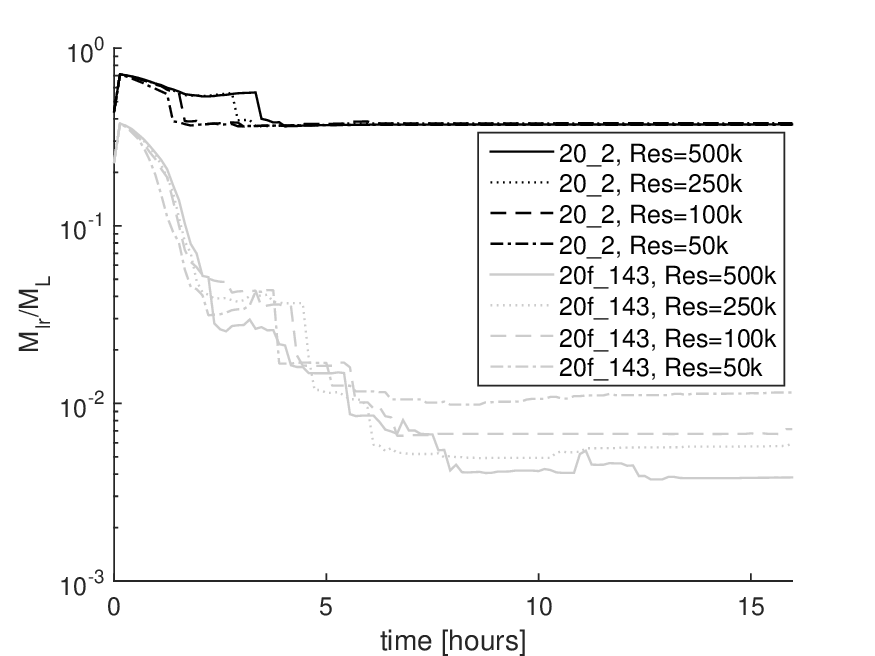}\label{fig:M_lr}\\
        \small (a) mass of largest fragment
    \end{tabular}
    \begin{tabular}[b]{c}
	   \includegraphics[width=0.507\textwidth]{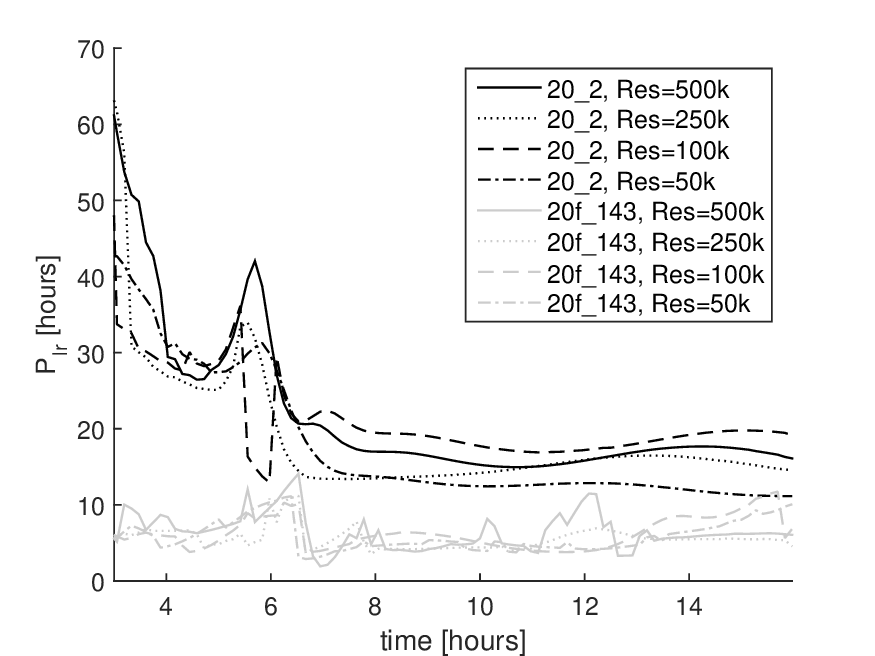}
        \label{fig:P_lr}\\
        \small (b) rotation period of largest fragment
    \end{tabular}
    
    \caption{Resolution convergence tests for scenarios 20\_2 (dark lines) and 20f\_143 (light lines). The mass of the largest fragment $M_{lr}$ (a) and its rotation period $P_{lr}$ (b) are shown as a function of time. Different resolutions are marked by various line textures shown in the legend.}
    \label{fig:Resolution}
\end{figure*}

As indicated in both scenarios, the simulation outcomes follow similar trajectories at all resolutions, and tend to converge to a similar value. As shown in panel (a), and as we see in other scenarios as well, masses evolve very similarly, including that of the largest fragment (and we have also checked other masses such as the total bound disc mass or ejected mass). Scenario 20f\_143 represents the most disruptive collision in our simulation suite. As such, the fragmentation is maximal, and given the small size of even the largest fragment, we do expect greater sensitivity to resolution (i.e., the ratio between the mass of the largest fragment and the minimum mass of an SPH particle increases and becomes more important when the resolution is lower). However, even in that case the largest fragment's mass does not change by more than a factor 2. The rotation period of the largest fragment is again similar in all cases, and can also change by up to a factor of 2.

The maximum or minimum values obtained after convergence do not seem to always correlate with the numerical resolution. Instead, their ordering appears to be different in every case. We conclude that different resolutions produce similar trends in the simulation data. Final values after convergence seem to be slightly affected by resolution, but the outcomes do not deviate significantly from each other. For our primary task of following-up on SPH simulations using N-body simulations, we consider the mapping of the fragment masses even at the lower resolutions sufficient for our purposes.

\label{lastpage}

\end{document}